\documentclass[letterpaper, usenatbib, fleqn, twocolumn]{mn2e} 
\usepackage{times}
\usepackage{graphicx}
\usepackage{amsfonts}
\usepackage{amsmath}
\usepackage{amssymb}
\usepackage{rotating} 
\usepackage{setspace}
\usepackage[ table ]{ xcolor }
\usepackage{pinlabel}
\DeclareMathAlphabet{\mathbf}{OML}{cmm}{b}{it}
\DeclareMathAlphabet{\mathbfsf}{OT1}{cmss}{bx}{n}

\newcommand\bq{\begin{equation}}
\newcommand\eq{\end{equation}}
\newcommand\bqs{\begin{equation*}}
\newcommand\eqs{\end{equation*}}

\newcommand\bqa{\begin{eqnarray}}
\newcommand\eqa{\end{eqnarray}}
\newcommand\bqas{\begin{eqnarray*}}
\newcommand\eqas{\end{eqnarray*}}

\newcommand{\be}{\begin{equation}}
\newcommand{\e}{\end{equation}}

\newcommand{\bear}{\begin{eqnarray}}
\newcommand{\ear}{\end{eqnarray}}

\def\u{{\bf U}}
\def\prd{PRD}

\def\u{{\vec U}}
\def\th{\vec{\theta}}

\def\jcap{J. Cosmology Astropart. Phys.}%
\def\aapr{A\&A~Rev.}%
\def\nar{New A Rev.}%
\def\nat{{ Nature}}
\def\apj{{ ApJ}}
\def\apjs{{ ApJS}}
\def\apjl{{ ApJ}}
\def\aap{{ A\&A}}
\def\aj{{ AJ}}

\def\mnras{{ MNRAS}}

\def\beq{\begin{equation}}
\def\eeq{\end{equation}}
\title[Prospects of observing a quasar HII region during the EoR region with redshifted 21cm] 
{Prospects of observing a quasar HII region during the Epoch of Reionization with redshifted 21cm}

\author[Datta et al.]{Kanan K. Datta$^{1}$
\thanks{e--mail: kdatt@astro.su.se},
Martina M. Friedrich$^{1}$, 
Garrelt Mellema$^{1}$, 
Ilian T. Iliev$^{2}$,
Paul R. Shapiro$^{3}$\\
$^1$Department of Astronomy \& Oskar Klein Centre, AlbaNova, Stockholm University, SE-106 91
Stockholm, Sweden \\
$^2$Astronomy Centre, Department of Physics \& Astronomy, Pevensey II Building, 
University of Sussex, Falmer, Brighton BN1 9QH\\
$^3$Department of Astronomy and Texas Cosmology Center, University of Texas, Austin, TX 78712, USA}

\date{\today}
\pubyear{2012} \volume{001} \pagerange{1}
\begin{document}
\pagerange{\pageref{firstpage}--\pageref{lastpage}}
\maketitle
\label{firstpage}
\begin{abstract}
We present a study of the impact of a bright quasar on the redshifted
21cm signal during the Epoch of Reionization (EoR). 
Using three different cosmological radiative transfer simulations, we
investigate if quasars are capable of substantially changing the size
and morphology of the H~II regions they are born in. We choose stellar
and quasar luminosities in a way that is favourable to seeing such an
effect. We find that even the most luminous of our quasar models is not
able to increase the size of its native H~II region substantially
beyond those of large H~II regions produced by clustered stellar
sources alone. However, the quasar H~II region is found to be 
more spherical. We next investigate the prospects of detecting such
H~II regions in the redshifted 21cm data from the Low Frequency Array
(LOFAR) by means of a
matched filter technique. We find that H~II regions with radii $\sim$
25 comoving Mpc or larger should have a sufficiently high detection
probability for 1200 hours of integration time. Although the matched
filter can in principle distinguish between more and less spherical
regions, we find that when including realistic system noise this
distinction can no longer be made. The strong foregrounds are found
not to pose a problem for the matched filter technique. We also
demonstrate that when the quasar position is known, the
redshifted 21cm data can still be used to set upper limits on the
ionizing photon rate of the quasar. If both the quasar 
position and its luminosity are known, the redshifted 21 cm data can set 
new constrains on quasar lifetimes.

\end{abstract}
\section{Introduction}
\label{intro}

The Epoch of Reionization (EoR) refers to the timespan between: (1) the formation of the first sources of light in the first halos that collapse under the influence of self-gravity in the highest density peaks in the neutral Universe after recombination; and (2) the time when most of the intergalactic hydrogen has become ionized by the radiation emitted from these sources of light. 
There are currently two key observations constraining this timespan. The optical depth, $\tau_{\mathrm es}$, due to Thompson scattering off free electrons, is measured by means of the polarization power spectrum of the cosmic microwave background (CMB). These measurements have constrained $\tau_\mathrm{es}$ to $0.088 \pm 0.015$, implying that an instantaneous reionization would have happened at $z=10.4\pm 1.2$ \citep{2010arXiv1001.4538K}. 
Quasar spectra obtained within the Sloan Digital Sky Survey (SDSS) indicate a low, but rapidly rising neutral fraction around redshift 6 \citep{2006AJ....132..117F, 2007AJ....134.2435W}. The combination of these two measurements suggests that the epoch of reionization extended over at least several redshift units. 

Radio telescopes capable of measuring at low frequencies, 
  (GMRT\footnote{Giant Metrewave Telescope,
  http://gmrt.ncra.tifr.res.in}, 21CMA\footnote{21 Centimeter Array,
  http://21cma.bao.ac.cn}, LOFAR\footnote{Low Frequency Array,
  http://www.lofar.org}, MWA\footnote{Murchison Widefield Array,
  http://www.mwatelescope.org}, PAPER\footnote{Precision Array to
  Probe the EoR, http://astro.berkeley.edu/\~{}dbacker/eor}) are either already existent or in the process of being build and aim to detect the signature of redshifted 21cm radiation from neutral hydrogen during the EoR. These measurements will provide completely new constraints on this early epoch of galaxy formation.

The sources of the ionizing photons during the EoR are potentially a mixture  of first galaxies, metal free POP~III stars and quasars. Although the contribution from quasars is believed to be small \citep[see e.g.][]{1996ApJS..102..191G, 2009JCAP...03..022L} due to their decreasing space density above redshift 2-3 \citep[e.g.][]{1995AJ....110...68S, 2000MNRAS.317.1014B, 2004ApJ...600L.119C, 2005BaltA..14..374G}, there are still doubts about the role of quasars in the EoR since secondary ionizations can yield several tens of ionizations per emitted high energy photon \citep{2009ApJ...703.2113V} and because galaxies alone might not be sufficient suppliers of ionizing photons \citep{2010MNRAS.409..855B}. 
\citet{2007MNRAS.374..627S} use a model for the quasar luminosity function together with a semi-analytic reionization description together with observations to set an upper limit to the quasar contribution to the ionizing background at $z\sim 5.8$ and find an upper limit of 14\%.
{Based on quasar survey data, \citet{2010AJ....139..906W} estimated the photon rate density from quasars to be between $20$ and $100$ times lower than the required rate to complete the reionization. \citet{2004ApJ...613..646D} and \citet{2006NewAR..50..204D} claim that the observed X-ray background can be used to set a limit on the contribution of power law type sources to global reionization.}

Apart from the observational limits on the number of quasars, it also remains unclear whether sufficiently massive black holes can form rapidly enough in order to create luminous quasars at redshifts relevant for hydrogen reionization. \citet{2010A&ARv..18..279V} suggests several different formation mechanisms for massive black holes. The recent {discovery} of a very luminous quasar at $z\sim7$ \citep{2011Natur.474..616M} shows however that there are in fact very massive black holes at such high redshifts. \citet{2011Natur.474..616M} estimate the mass of the central black hole to be of the order of $ 10^9 M_{\odot}$ based on Mg~II line width measurements and the luminosity at $\lambda=300$~nm using the method described in \citet{2009ApJ...699..800V}.  

{This highest redshift quasar was detected as part of the UKIRT Infrared Deep Sky Survey (UKIDSS) which is expected to find $\sim 10$ more quasars at $z>6.5$ \citep{2007MNRAS.379.1599L}. Further surveys for high redshift quasars are being performed with ESO's VISTA telescope \citep{2010Msngr.139....2E}, the Panoramic Survey Telescope and Rapid Response System 1 \citep[PAN-STARRS][]{2002SPIE.4836..154K, 2011arXiv1109.6241M} and the Canadian-France-Hawaii Telescope (the CFHQSIR program). These surveys will add to the already existing list of high redshift quasars established by the Sloan Digital Sky Survey \citep[SDSS][]{2001AJ....122.2833F, 2006AJ....132..117F, 2008AJ....135.1057J} and the Canadian-France High Redshift QSO survey \citep[CFHQS][]{2007AJ....134.2435W, 2010AJ....139..906W}. All of the above efforts concentrate on the redshift interval 6 to 7. To map out the quasar population for $z\la 8$ a space mission such as ESA's EUCLID may be needed \citep{2012MNRAS.420.1764R}.
}

While the quasar contribution to the ionizing photon budget is a matter of ongoing debate, it is generally accepted that quasars (including the so called {mini- and micro-quasars) contribute substantially} to the (global) heating of the neutral IGM before large scale ionization. Several EoR simulations \citep[e.g.][]{2010A&A...523A...4B, 2007MNRAS.375.1269Z} include a {quasar-like} population to investigate effects of heating. There are other groups that include a mixture of sources to test the large scale effect of quasars on the morphology of H~II regions, for example \citet{2009MNRAS.399.1877G} who include a quasar population in a semi-analytic fashion and investigate their effect on the 21cm power spectrum.

A different approach is taken by \citet{2007MNRAS.376L..34M}, who include a single quasar in a reionization simulation to generate Ly$\alpha$ absorption spectra. They concentrate on the very late stages of reionization, around $z\geq 6$, and assume an IGM in photoionization equilibrium with some uniform ionizing background. Their main goal is to investigate the reliability of measurements of the average ionization state of the IGM from estimates of near-zone sizes from quasar spectra. They come to a similar conclusion as  \citet{2007ApJ...670...39L} who use a semi-numerical approach to model quasar spectra from a quasar in a patchy ionized IGM, namely that the effect of small scale inhomogeneities in the IGM is larger than the effect of the global ionization fraction and therefore conclude that  Ly$\alpha$ absorption spectra of high redshift quasars give only limited information about the global ionization fraction.
Additionally, \citet{2007MNRAS.380L..30A} point out the importance of the pre-existence of stellar H~II regions at the locations of quasars when they begin to shine. Neglecting the stellar H~II region will
will lead to an overestimate of the calculated average surrounding ionization fraction. 

A similar approach of concentrating on a single quasar has been taken recently by \citet{2011arXiv1111.6354M} and earlier for example by \citet{2008MNRAS.386.1683G}. 
While the latter investigate the effect of quasar H~II regions on mock 21cm line-of-sight spectra to extract information about the global ionization fraction, the former use a filter technique described first in \citet{2007MNRAS.382..809D} to investigate quasar properties from 21cm observations. 
Our paper has a similar goal, namely to study the imprint of a single bright quasar on the morphology of H~II regions and its observability using 21cm observations.
The main difference between these previous approaches and ours is that they use a semi-numerical method for generating the ionization fraction field including the quasar while we perform full radiative transfer in a hydrogen/helium IGM. %

This paper tries to answer the following questions: Does a quasar change the size and the morphology of an H~II region in a distinct manner? Can LOFAR in combination with the matched filter approach be used to detect individual H~II regions? Can we use the matched filter technique in a blind search (i.e. without prior knowledge of a quasar location) to identify an H~II region as a quasar H~II region? If we have prior knowledge about the quasar position, can we use the matched filter technique to extract quasar properties from the H~II region size estimated?
To answer these questions, we test three cases of detectability of a single quasar H~II region in a patchy ionized environment.

The structure of the paper is as follows. In Sect. \ref{sec:sims} we describe our simulations. In Sect. \ref{sec:results1} we compare the ionization fraction fields and differential brightness temperature maps of the simulations without a quasar to the ones with a quasar and make some estimate on the expected quasar H~II region sizes based on photon counts. We try to asses the three dimensional shape of the H~II regions. Sect. \ref{sec:visibilities} describes how to simulate the radio-interferometric visibilities from the redshifted 21cm line radiation. Sect. \ref{sec:matched_filter} summarizes the matched filter method which we use to detect the quasar H~II regions in Sect. \ref{sec:results2}. Sect. \ref{sec:foregrounds} discusses the role of foregrounds. In Sect. \ref{sec:summary} we summarize our findings. 


\section{Simulations}
\label{sec:sims}

The general outline of our simulations that we perform to test the detectability of quasar H~II regions is the following: \\
(a) We start with the results of an already existing large scale hydrogen only radiative transfer reionization simulation with stellar-type sources. \\
(b) In this simulation, we select a redshift $z_{\rm on}$ where the global ionization fraction is small enough that the individual H~II regions of galaxy clusters, although already partly connecting, can still be recognized as individual H~II regions.  \\
(c) We add helium, initialized to have the same ionization state as hydrogen some millions of years before the quasar turn on and run the simulation until we reach $z_{\rm on}$ in order to have better initial conditions for helium when including the quasar.\\
(d) At $z_{\rm on}$, we select the most massive halo. Here we include one additional source, a quasar with an ionizing photon rate that is constant in time and which is chosen on the basis of the host-halo mass. \\
(e) We run the simulation for another 23 Myr with all stellar sources and the quasar, following the ionization of hydrogen and helium. \\
In the following, we describe these steps in more detail. 

\subsection{Simulations without a quasar}
The simulation methodology (with stellar type sources only) has been described in detail in \citet{2006MNRAS.369.1625I}, \citet{methodpaper} and \citet{2007MNRAS.376..534I}. Here we only summarize some important aspects. 
The density fields through which we trace the ionizing radiation and the sources that emit this ionizing radiation are extracted from cosmological N-body simulations of structure formation. For these N-body simulation, the \textsc{CubeP$^3$M} code was used. This code was developed from the PMFAST code \citep{2005NewA...10..393M}, see \citet{2008arXiv0806.2887I} for a short description of the \textsc{CubeP$^3$M} code. It uses particle-particle interactions at sub grid distances and particle-mesh interactions for larger distances. We use the results of two such simulations: one with a volume of (163~cMpc)$^3$ that has $3072^3$ particles and a mesh size of $6144^3$ cells; and one with a volume of (607~cMpc)$^3$ that has $5488^3$ particles and a mesh size of $10976^3$ cells.\footnote{As is described in detail below, we only resimulate a sub-volume of the latter.} This implies particle masses of $5.5\times 10^6$~M$_{\odot}$ and $5.0\times 10^7~M_{\odot}$, respectively. 
The first set of parameters guarantees a minimum resolved halo mass of $\sim 10^8$~M$_\odot$ which is approximately the minimum mass of halos able to cool by atomic hydrogen cooling. The second set of parameters guarantees a minimum resolved halo mass of $\sim 10^9$~M$_\odot$. 

All N-body simulations use the cosmological parameters for a flat $\Lambda$CDM universe with $(\Omega_m,\Omega_b,h,n,\sigma_8) = (0.27,0.044,0.7,0.96,0.8)$,  based on the five year WMAP results \citep{2009ApJS..180..330K}. For creating the halo list, a spherical overdensity halo finder was used.

The density is assigned to a coarser mesh ($256^3$ and $504^3$, respectively) by smoothing the dark matter particle distribution using an SPH kernel function. To convert the dark matter density into a baryon density, we assume that the gas distribution follows the dark matter. This is valid on the scales of the radiative transfer cells (0.64 and 1.2 comoving Mpc, respectively) since the local Jeans length at the mean density of the IGM is much smaller than that. For the radiative transfer simulations, we use the short characteristic ray tracing method \textsc{C$^2$-Ray} \citep{methodpaper,2012MNRAS.tmp.2385F} on this coarser mesh.

Each halo of mass $M$ is assigned an ionizing photon luminosity
\beq
\dot{N}_\gamma=g_\gamma\frac{M\Omega_{\rm b}}{10 \, \rm{Myr} \, \Omega_m m_p}\,,
\eeq
where $\dot{N}_\gamma$ is the number of ionizing photons emitted per Myr, $M$ is the halo mass, and $m_p$ is the proton mass.  Halos are assigned different luminosities according to whether their mass is above 
or below 
$10^9 M_\odot$ (but above $10^8 M_\odot$). For the second simulation, we use a recipe calibrated against the 163 cMpc simulation, to include sources in halos with masses $10^8 M_{\odot} \leq M \leq 10^9 M_{\odot}$.
Sources in low mass halos are suppressed if the ionization fraction of their cell is larger than 10\%. In the simulations we use here, sources in high mass halos have an efficiency $g_\gamma=1.7$ and  sources in low mass halos have an efficiency $g_\gamma=8.7$. 

These are rather low efficiencies for the stellar sources. This
combined with our prescription for the quasar luminosity (see Sect. \ref{subsec:qso}) means that
the quasar will outshine the stellar sources in its host halo by a factor of a few, see Table \ref{table:summary2} for a summary of stellar and quasar photon rates. This is necessary for the quasar to have any impact on the
neutral hydrogen distribution in its environment as otherwise the stellar sources
in neighbouring halos would always dominate the photon budget in the H~II region
surrounding this high density region. 

The first simulation used here, 163Mpc\_g1.7\_8.7S is one simulation of a set of simulations presented and analysed in \citet{2011arXiv1107.4772I}. It has a simulation volume of (163 cMpc)$^3$.
The second simulation, 607Mpc\_g1.7\_8.7S has a considerably larger volume. It will be analysed and described in detail in Iliev et al., in prep. 

The 163 cMpc hydrogen only simulation without the QSO reaches global ionization fractions $\left<x_{\rm m}\right>$=(0.1, 0.5, 0.7, 0.9 and 0.99) at redshifts $z=$(9.9, 7.6, 7.2, 6.9 and 6.7) \citep[see][]{2011arXiv1107.4772I}. The 607 cMpc simulation has currently not reached more than  $\left<x_{\rm m}\right>=0.7$, the redshifts  for global ionization fractions (0.1, 0.5, 0.7) are (9.6, 7.5, 7.1). Although the two simulations were performed using the same source properties, it is visible that the ionization history is slightly delayed in the larger simulation volume. This is most probably due to the way in which the suppressible sources are included, which results globally in a higher suppressed fraction.
The 163~cMpc simulation has an electron Thompson scattering optical depth $\tau=0.058$.

\begin{table}
\caption{Summary of important quantities of the three simulated cases.}

\resizebox{\columnwidth}{!}{
\begin{tabular}{l  c c c c c  }

\hline
 & resolution	& box (sub-box)  &  $z$(QSO on) &  $z$(QSO on  & $z$(QSO on  \\
 &                & size           &            &    +11.5 Myr)  &   +23.0 Myr) \\
                  \hline \\[-6pt]
early QSO & 0.64 cMpc & 163 cMpc & 8.636 &  8.515 & 8.397 \\
late QSO  & 0.64 cMpc & 163 cMpc & 7.760 &  7.664 & 7.570 \\
large box        & 1.2 cMpc & 607 (242) cMpc & 7.760 & 7.664 & 7.570 \\[1pt]
\hline

\end{tabular} }
\label{table:nonlin}
\end{table}
\subsection{Adding a quasar to the simulations}
\label{subsec:qso}
\begin{figure}
  \centering
 \includegraphics[clip,width=8.5cm]{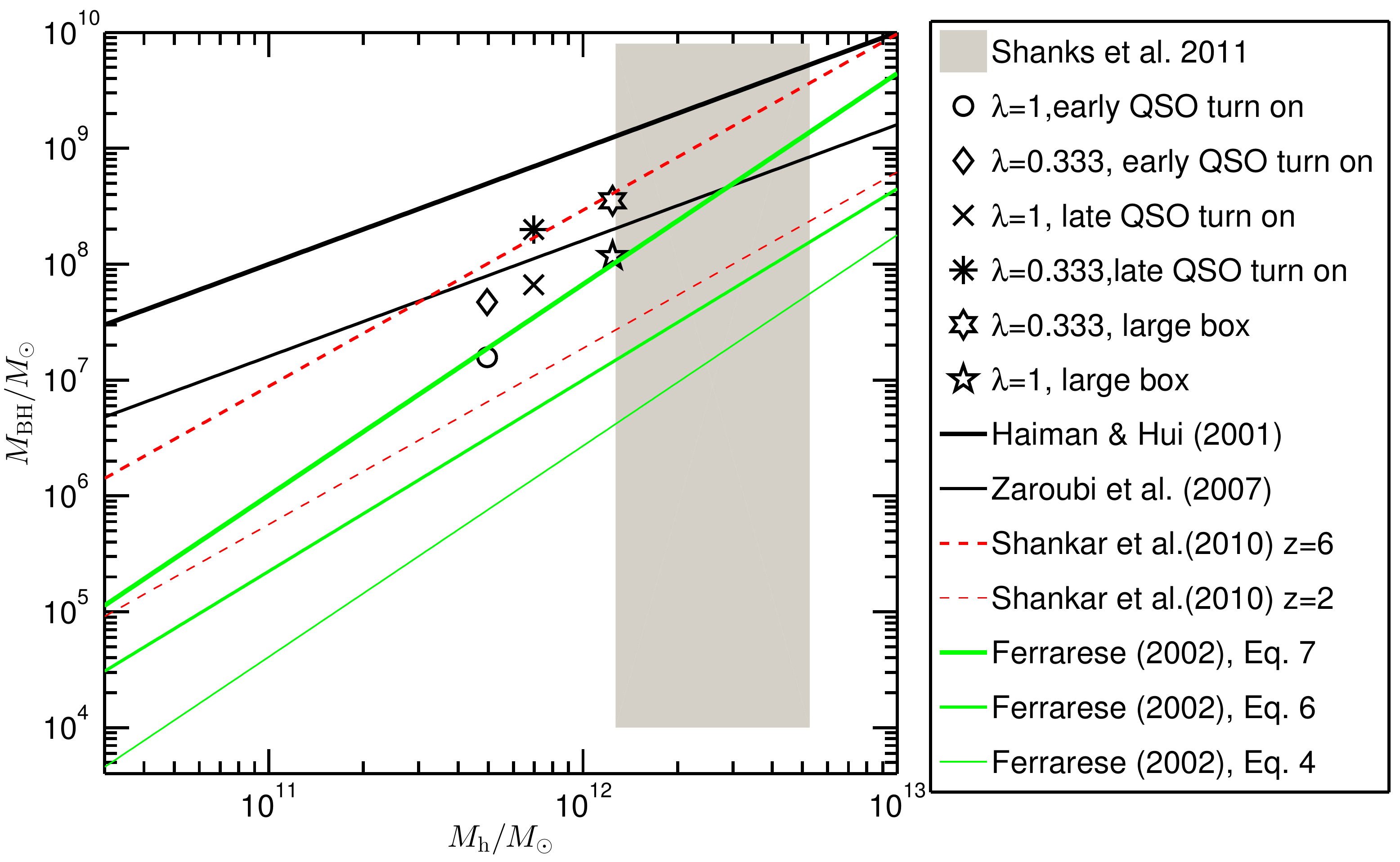}
\caption{Different analytic estimates for the halo mass $M_{\rm h}$ - black hole mass $M_{\rm BH}$ relation. Note the  large spread. The \citet{2010ApJ...718..231S} curves are produced using their fiducial parameters. \citet{2011MNRAS.416..650S} and \citet{2005MNRAS.356..415C} suggest a  narrow host halo mass range for quasars, indicated in the figure by the grey rectangular shaped region. Included in this plot are the quasars from our simulations. We include two possible black hole masses for each quasar implementation, one assuming $\lambda=1$ and one assuming $\lambda=0.333$, which both produce the same luminosity. \citet{2012arXiv1201.5383D} find that the distribution of Eddington ratios peaks between $\lambda= 0.1 \-- 0.2$ for black hole masses between $M_{\rm BH} = (10^7 \-- 10^8) M_{\odot}$. The halo mass for the six points corresponds to the mass of the most massive halo for the indicated cases. }
\label{fig:MbMBH_scatter}
\end{figure}

For adding a quasar, the next step to take is to identify a suitable redshift. 
Observationally, the highest redshift quasar has been detected at $z=7.085$ \citep{2011Natur.474..616M}. 
However, we choose higher redshifts ($z = 8.636$ and $z = 7.760$ for the quasar turn on) because of the lower global (i.e. mass averaged)
ionized fractions $\left<x_{\rm m}\right>$ at those redshifts (0.25 and 0.45) in our simulation. At these epochs the size distribution
of H~II regions still allows them to be rather isolated and thus makes it possible for a quasar to
modify the H~II region that it finds itself in. As was shown for example by \citet{2011MNRAS.413.1353F} at higher global ionized fractions most H~II regions have become connected
and the topology of the IGM is rather complex. \citet{2008MNRAS.386.1683G} also already concluded that beyond $\left<x_{\rm m}\right>\sim 0.7$ the signature of the quasar
is difficult to recognize in the neutral hydrogen distribution. 

As was explained in \citet{2011MNRAS.413.1353F} a reionization simulation of a given volume
can only accommodate a limited number of H~II regions of a given size at a given global
ionized fraction. This means for our case that the (163~cMpc)$^3$ volume may possibly only
accommodate one large H~II region from which we would then conclude that the H~II region
containing the quasar could be recognized as being the largest one. We therefore also
use the larger volume of (607 cMpc)$^3$ which allows us to address the question whether there
are H~II regions powered by stellar sources alone which could obtain a similar size
as the quasar containing H~II region. An additional advantage of using this larger
volume is that the most massive halo corresponds to a higher $\sigma$--peak in the
density field and is therefore more massive.

However, because of the larger volume, our procedure for adding the quasar to the 607 cMpc box
differs a little from that for the 163 cMpc one. Firstly, the radiative transfer
of ionizing photons in the larger volume is limited by a mean free path set to 60 cMpc,
roughly corresponding to the mean free path of ionizing photons due to Lyman Limit Systems
at $z\sim 6$ \citep{2010ApJ...721.1448S}. We therefore only need to resimulate part of the
volume to assess the impact of the quasar as photons further away than 60 cMpc from the region of interest will not reach this region.
In order to insert the quasar region back into the whole volume we actually need to include
a buffer zone of the same extend around our resimulated region, so the ``zoomed'' region which
we resimulate with a quasar has a size of (242 cMpc)$^3$ centred on the quasar and consists of
$201^3$ radiative transfer cells. This procedure introduces the following differences
compared to the quasar simulations in the smaller volume: 

\noindent (1) By taking a sub-box, the average density in the box is no longer equal to the average density of the universe. In fact, with our choice of size, the sub-box is slightly less dense than the average density of the universe. However, this is only a minor deviation, the average density in the sub-box is only 0.25\% lower than the average density in the whole simulation volume. 

\noindent(2) The spatial resolution of this simulation is worse than for the (163 cMpc)$^3$ simulation, 1.2 cMpc instead of 0.64 cMpc.

\noindent(3) Since we take a sub-volume of a larger simulation, we use open boundary conditions. When inserting the subvolume back into the full volume we discard the outer buffer zone. 

The basic simulations described above, 163Mpc\_g1.7\_8.7S and 607Mpc\_g1.7\_8.7S do not include helium. This is equivalent to assuming that helium is singly ionized everywhere where hydrogen is ionized and that double ionization of helium is negligible. This is an assumption frequently made in EoR simulations including stellar type sources \citep[e.g.][]{2006MNRAS.369.1625I, 2006MNRAS.372..679M, S8, S9, 2010ApJ...724..244A}. As was demonstrated in \citet{2012MNRAS.tmp.2385F}, this is a valid assumption in the case of stellar type sources. Since quasars emit a substantial amount of their photon output at energies that are sufficient to doubly ionize helium, and since the ionization cross-section of helium is roughly an order of magnitude larger than that of hydrogen, it is essential to include helium in our simulations containing a quasar. In order to do so, we initialize the helium ionization fractions 11.5 Myr before the quasar turn on for each of the three cases to correspond to that of hydrogen. He~III is assumed to be absent. We showed in \citet{2012MNRAS.tmp.2385F} that the He~I fraction field  corresponds approximately to the H~I fraction field. We evolve the so generated helium ionization fraction field together with the hydrogen for 11.5 Myr before adding the quasar source to get better initial conditions for the helium fraction fields. The corresponding redshifts for adding helium are 8.762 for the early quasar turn of simulation 163Mpc\_g1.7\_8.7S and 7.859 for the late quasar turn on of simulation 163Mpc\_g1.7\_8.7S and simulation  607Mpc\_g1.7\_8.7S. In Table \ref{table:nonlin} we summarize all relevant quantities for all three (\textbf{early} and \textbf{late} QSO turn on of simulation 163Mpc\_g1.7\_8.7S and the \textbf{large box} simulation 607Mpc\_g1.7\_8.7S) studied cases.

The general idea for the quasar model is to connect the quasar luminosity to the host halo mass. This can be done in three steps: \\
(1) The quasar luminosity $L_{\rm QSO}$ can be connected to the black hole mass $M_{\rm BH}$ \citep[reverberation mapping shows that there is a correlation between black hole mass and quasar luminosity, ][]{2008ApJ...680..169S} \\
(2) $M_{\rm BH}$ can be connected to the mass of galaxy bulges $M_{\rm B}$.  \citep[Maggorian type relation, ][]{1998AJ....115.2285M,2003ApJ...589L..21M,2004ApJ...604L..89H,2007ApJ...663...53T,2012ApJ...746..113G}. \\
(3) $M_{\rm B}$ can be connected to the total mass of the halo $M_{\rm h}$ \citep{2002ApJ...578...90F}. \\
Combining these relations, there is a big scatter for the relation halo mass to black hole mass, see Fig. \ref{fig:MbMBH_scatter}. \citet{2011MNRAS.416..650S} argue that such a general relation does not exist. Instead, they suggest that there is a typical halo mass range for active quasars. Also \citet{2011Natur.469..377K} argue that there is in general no correlation between the halo mass of arbitrary galaxies and the central black hole.
However, due to the lack of better observational or theoretical restrictions, we base our quasar luminosity on the host halo mass.

Since we aim for a luminous quasar, we orientate us towards the upper limit of the conversion factor between halo mass and black hole mass, see Fig. \ref{fig:MbMBH_scatter}. The mass of the black hole can be converted into a luminosity by assuming an Eddington limited accreation
\bq
L_{\rm QSO}= \lambda L_{\rm Edd}= \lambda \frac{4 \pi M_{\rm BH} m_{\rm p} c}{\sigma_{\rm T}} \approx \lambda \, 1.3 \times 10^{31}  \frac{M_{\rm BH}}{M_{\odot}}\rm{W} \, .
\eq
Here, $G$ is the gravitational constant, $M_{\rm BH}$ the mass of the black hole, $m_{\rm p}$ the proton mass, $c$ the speed of light and $\sigma_{\rm T}$ the Thompson scattering cross section. $\lambda$ is the Eddington efficiency parameter.
Although Eddington efficiencies $\lambda$ larger than 1 are in principle possible \citep[see for example][ for
simulations assuming a thin disc]{2007IAUS..238..153H}, observations show that most quasars are below the Eddington
limit \citep[see for example figure 1 in][]{2010MNRAS.402.2637S}. Also hydrodynamical simulations indicate that most quasars shine with luminosities below the Eddington luminosity \citep{2012arXiv1201.5383D}. In Fig. \ref{fig:MbMBH_scatter}, we include the
corresponding black hole masses for two assumed values for $\lambda$, to give an impression of how our choice of
parameters compares with other models. The important parameter for us is the total luminosity and not the
black hole mass, so there is a degeneracy between $\lambda$ and $M_{\rm BH}$. Therefore, the six points in the plot only correspond to three different simulations with three different quasar luminosities.  The conversion factor between halo mass and quasar luminosity, $L_{\rm QSO} = \mathcal{K} M_{\rm h}$ for the early quasar turn on is $\mathcal{K}=4.3 \times
10^{26} {\, \rm W} M_{\odot}^{-1}$ and for the late quasar turn on in the 163~cMpc and 607~cMpc simulations  $\mathcal{K}=1.3 \times 10^{27} {\, \rm W}
M_{\odot}^{-1}$. The most massive halo in the three cases (early turn on, late turn on and large box) are $4.9 \times 10^{11} M_{\odot}, \, 6.9 \times 10^{11} M_{\odot}$ and $1.2 \times 10^{12} M_{\odot}$, respectively.
With our choice of power law index ($\alpha= -1.5, \, L(\nu)\propto \nu^{\alpha}$) and with our high energy
cut off at 5440 eV, this corresponds to  $ 3.3\times 10^{55}$~s$^{-1}$, $1.4 \times 10^{56}$~s$^{-1}$ and $2.4 \times 10^{56}$~s$^{-1}$ ionizing photons per second for the early and late quasar turn on of the 163 cMpc simulations and the 607 cMpc simulation, respectively. We summarize these numbers in Table \ref{table:summary2}. The recently observed $z=7.085$ quasar is estimated to have an ionizing photon rate of $1.3 \times 10^{57}$~s$^{-1}$ \citep[][]{2011Natur.474..616M}. The lower redshift ($z\sim 6$) quasar that \citet{2007MNRAS.376L..34M} investigate in their simulations has an ionizing photon rate of $5.2 \times 10^{56}$ s$^{-1}$ and the quasar included in the simulations of \citet{2011arXiv1111.6354M} has an ionizing photon rate of $8.97 \times 10^{56}$ s$^{-1}$. Although our most luminous quasar has a lower ionizing photon rate by a factor of a few compared to these other studies, the resulting H~II regions have comparable sizes since we consider a longer quasar lifetime. The reason we have a slightly less luminous quasar is that our most massive halo in our simulation is less massive than the most massive halo in, for example, \citet{2007MNRAS.376L..34M} since we are considering higher redshifts. 

\citet{2012A&A...537L...8C} observe a $z=2.4$ quasar and find  that star formation in the host halo is quenched. We do not include negative feedback from the quasar on the star formation other than the usual suppression of low-mass sources in ionized regions. However, since the quasar dominates the photon output during its on-time, see Table \ref{table:summary2}, and since \citet{2008MNRAS.391...63I}  found that the contribution of the largest halo  to the total photon-budget in a jointly produced H~II region is small compared to the sum of photons from the clustered fainter sources contributing to the H~II region, we argue that negative feedback on the star formation would not change our conclusions significantly. Another important assumption is that the quasar emits radiation equally in all directions. Although this is in line with previous studies of this kind \citep{2008MNRAS.386.1683G,2011arXiv1111.6354M}, real quasars may emit non-isotropically, which would change the geometry of their region of influence. Considering this complication is beyond the scope of this work.


\section{Results from the simulations}
\label{sec:results1}

\begin{table}
\caption{Summary of important results for the three cases; The stellar ionizing photon rates $\langle \dot{N}_{\gamma}^{*}\rangle$ are averages over the stellar photons emitted in the quasar area of influence during the quasar-on time, $N_{\gamma}^*$ and $N_{\gamma}^{\rm qso}$ denote the total number of stellar and quasar photons emitted in the region of influence of the quasar over its entire history, respectively. $M_{\rm h, max}$ denotes the mass of the most massive halo in each simulation volume.}
\resizebox{\columnwidth}{!}{
%
\begin{tabular}{l  c c c c c }
\hline \\[-6pt]
 & $\langle \dot{N}_{\gamma}^{*}\rangle$/s & $\dot{N}_{\gamma}^{\rm qso}$/s & $M_{\rm h,max}/M_{\odot}$ & 
 								$N_{\gamma}^*$ & $N_{\gamma}^{\rm qso}$ \\[6pt]
                  \hline \\[-5pt]
early QSO & $4.5 \times 10^{54}$ & $3.3 \times 10^{55}$ & $4.9 \times 10^{11}$ & $ 2.5 \times 10^{70}$ & $2.5 \times 10^{70}$ \\[1pt]
late QSO    & $1.7 \times 10^{55}$ & $1.4 \times 10^{56}$ & $6.9 \times 10^{11}$ & $9.8 \times 10^{70}$ &$1.0 \times 10^{71}$\\[1pt]
large box         & $4.3 \times 10^{55}$ & $2.4 \times 10^{56}$ & $1.2 \times 10^{12}$ & $2.2 \times 10^{71}$ & $1.7 \times 10^{71}$\\[1pt]
\hline
\end{tabular}
} 
\label{table:summary2}
\end{table}

\begin{figure*}
  \centering
\labellist
\small\hair 2pt
\pinlabel $\textbf{(a)}$ at 20 590 
\pinlabel \rotatebox{90}{$\xleftarrow{\hspace*{.9cm}} \textbf{163 cMpc}\xrightarrow{\hspace*{.9cm}}$} at -4 431
\endlabellist  
\includegraphics[width=8.4cm]{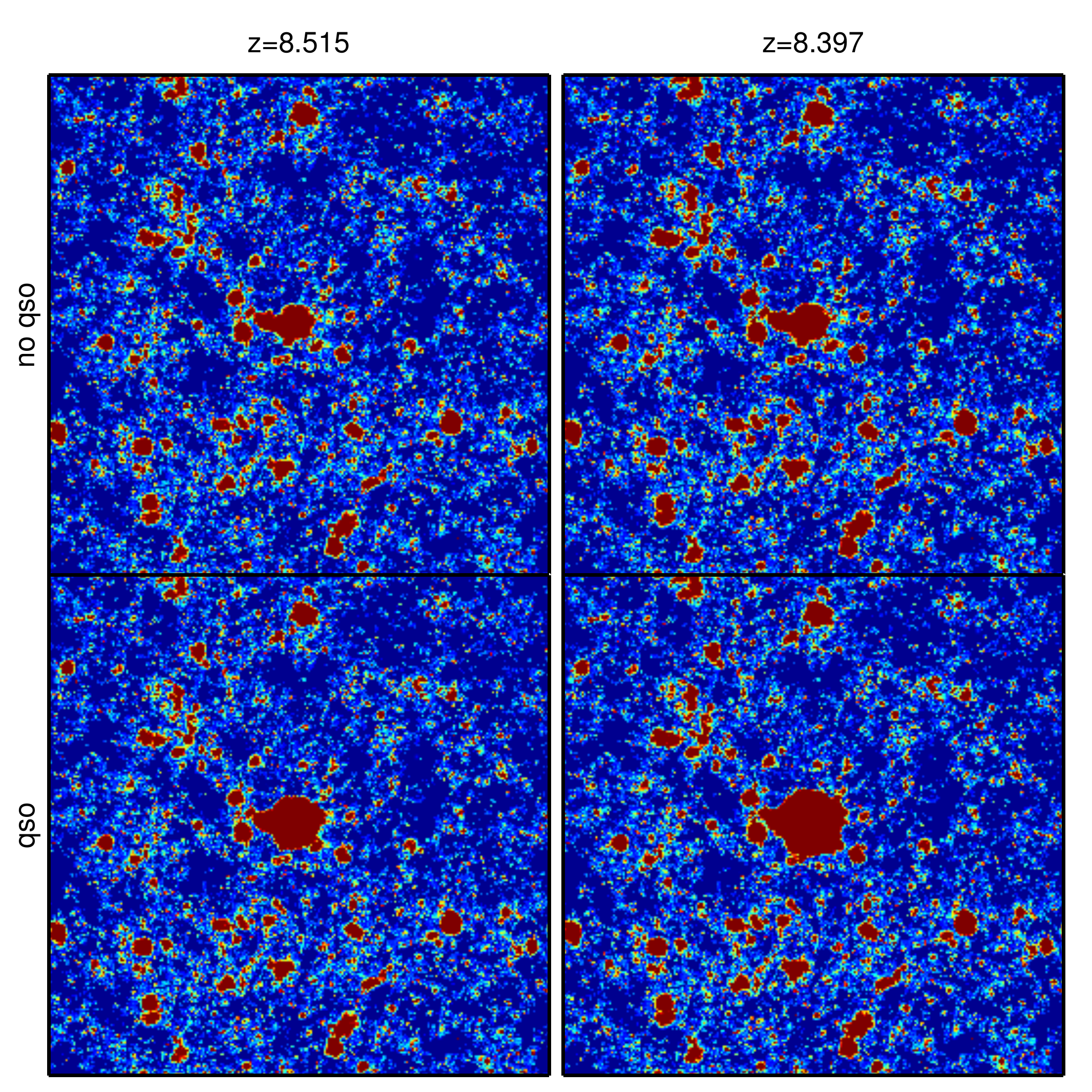}  
\labellist
\small\hair 2pt
\pinlabel $\textbf{(b)}$ at 20 590 
\pinlabel \rotatebox{90}{$\xleftarrow{\hspace*{.9cm}} \textbf{163 cMpc}\xrightarrow{\hspace*{.9cm}}$} at 624 431
\endlabellist 
\includegraphics[width=8.4cm]{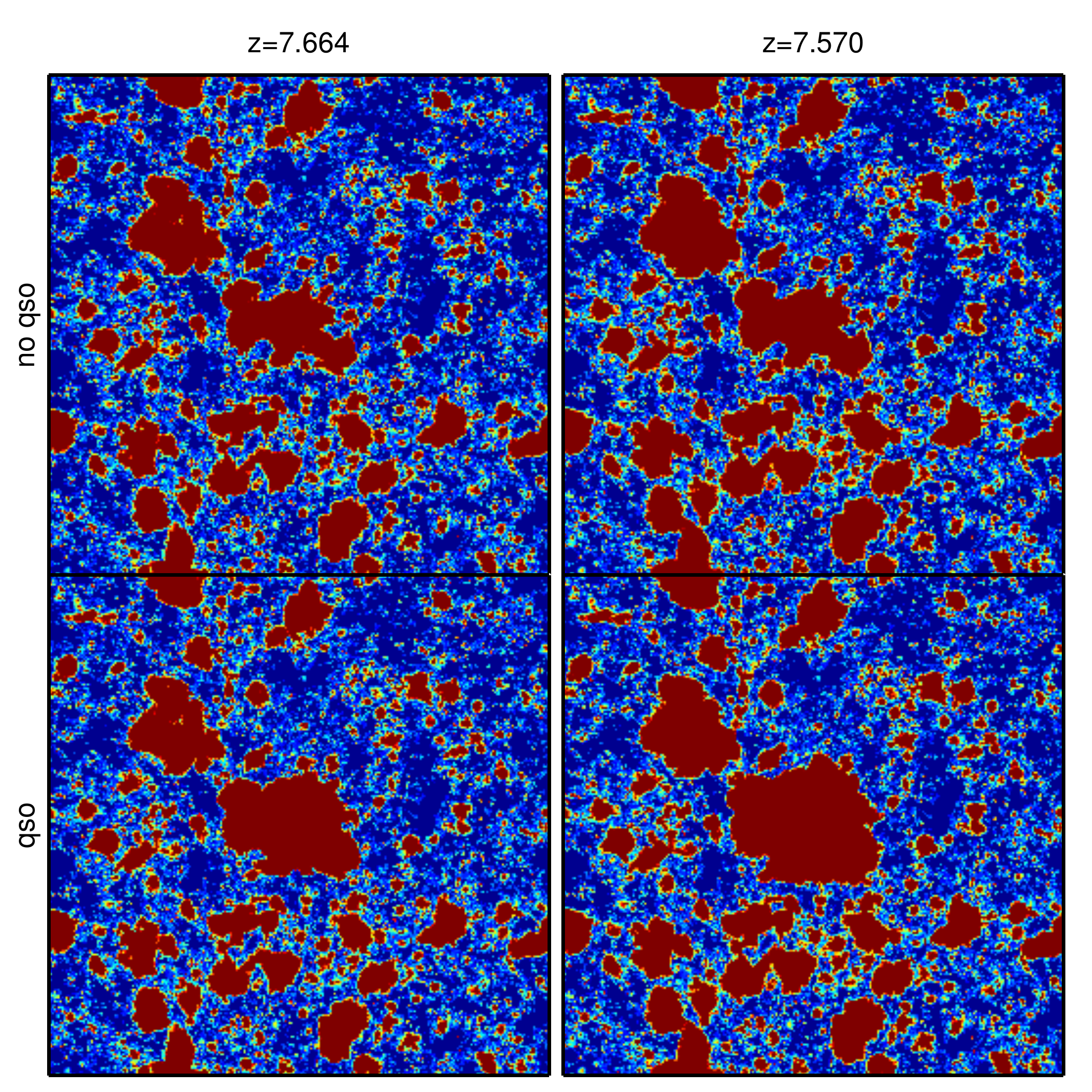}  
\labellist
\small\hair 2pt
\pinlabel $\textbf{(c)}$ at 20 455 
\pinlabel \rotatebox{90}{$\xleftarrow{\hspace*{2.7cm}}\textbf{607 cMpc} \xrightarrow{\hspace*{2.7cm}}$} at 1 223
\endlabellist
\includegraphics[width=8.4cm]{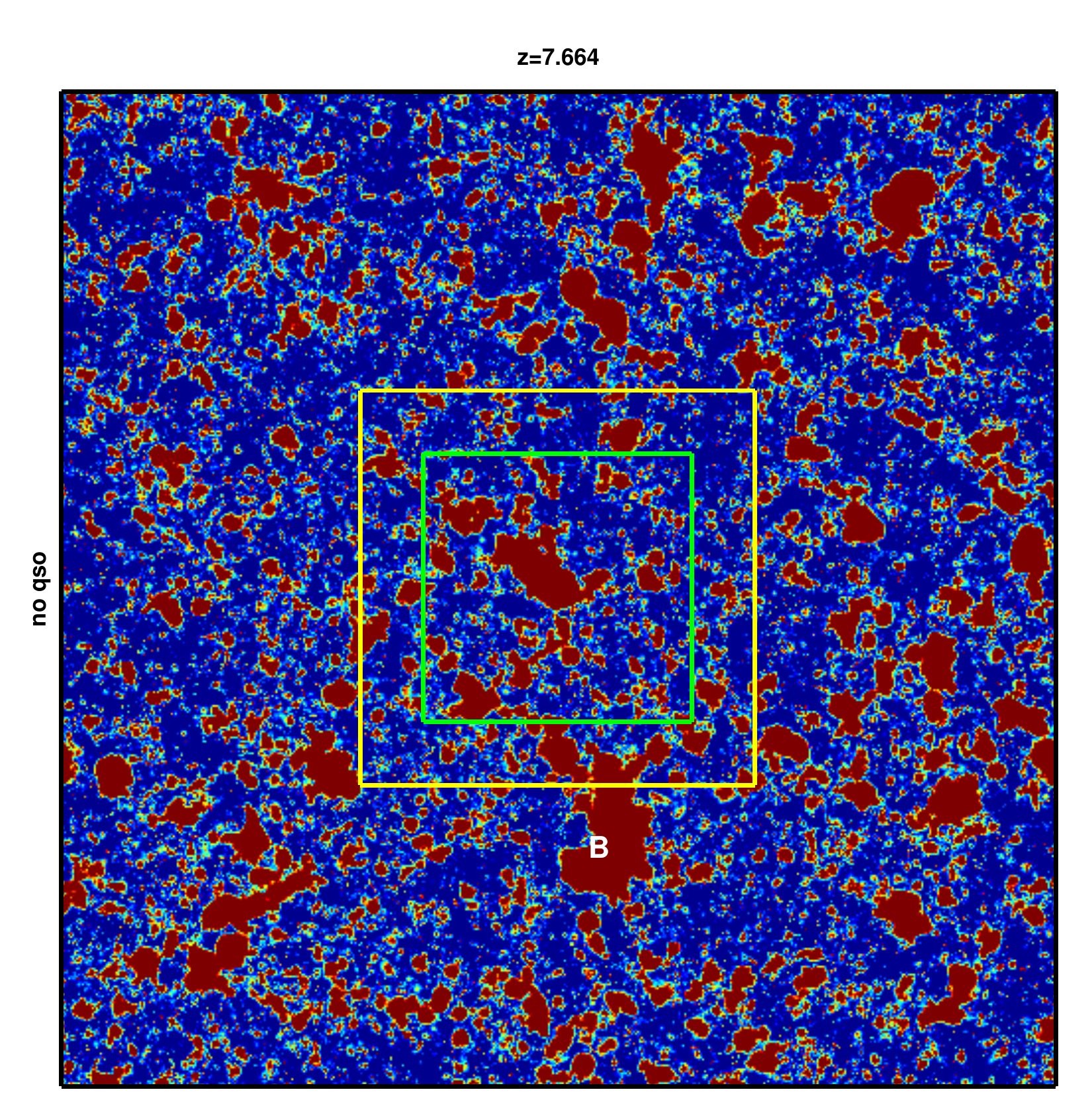}
\labellist
\small\hair 2pt
\pinlabel $\textbf{(d)}$ at 20 590 
\pinlabel \rotatebox{90}{$\xleftarrow{\hspace*{.9cm}} \textbf{163 cMpc}\xrightarrow{\hspace*{.9cm}}$} at 624 431
\endlabellist
\includegraphics[width=8.4cm]{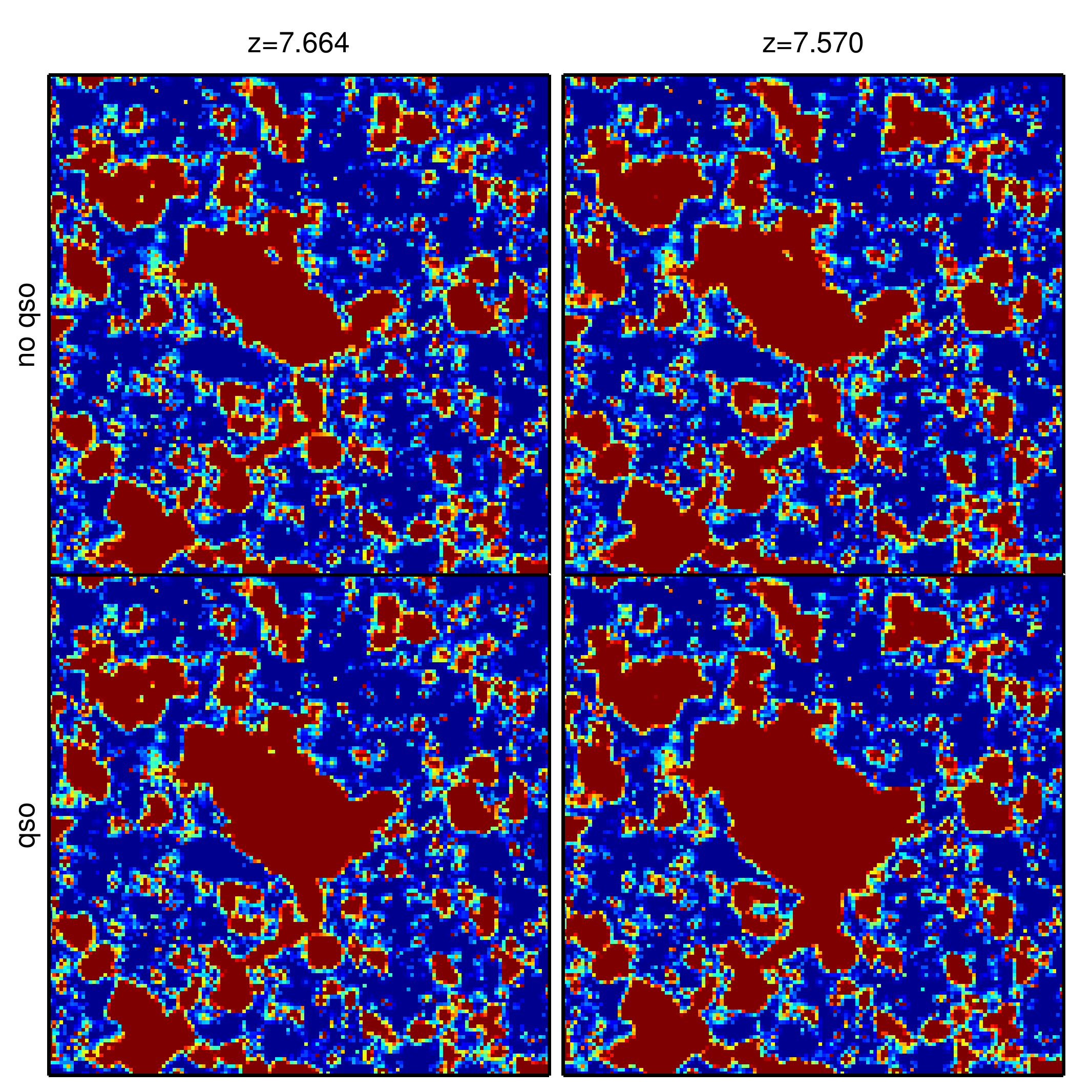}
\caption{Slices of hydrogen ionization fraction, blue corresponds to fully neutral, red to fully ionized. The upper left four slices (panel a) correspond to the early quasar turn on in the 163Mpc\_g1.7\_8.7S simulation, the upper right four slices (panel b) correspond to the late quasar turn on in this simulation. The lower right four slices (panel d) correspond to a sub-region of the same size (163 Mpc) of the 607Mpc\_g1.7\_8.7S simulation. In each panel, the upper row correspond to a simulation without quasar and the lower row to the same simulation including a quasar in the centre. For completeness, we also show a slice of the whole 607Mpc\_g1.7\_8.7S (panel c) simulation. The yellow square indicates the re-simulated part and the green square indicates the sub-region show in panel d.}
\label{fig:slices}
\end{figure*}
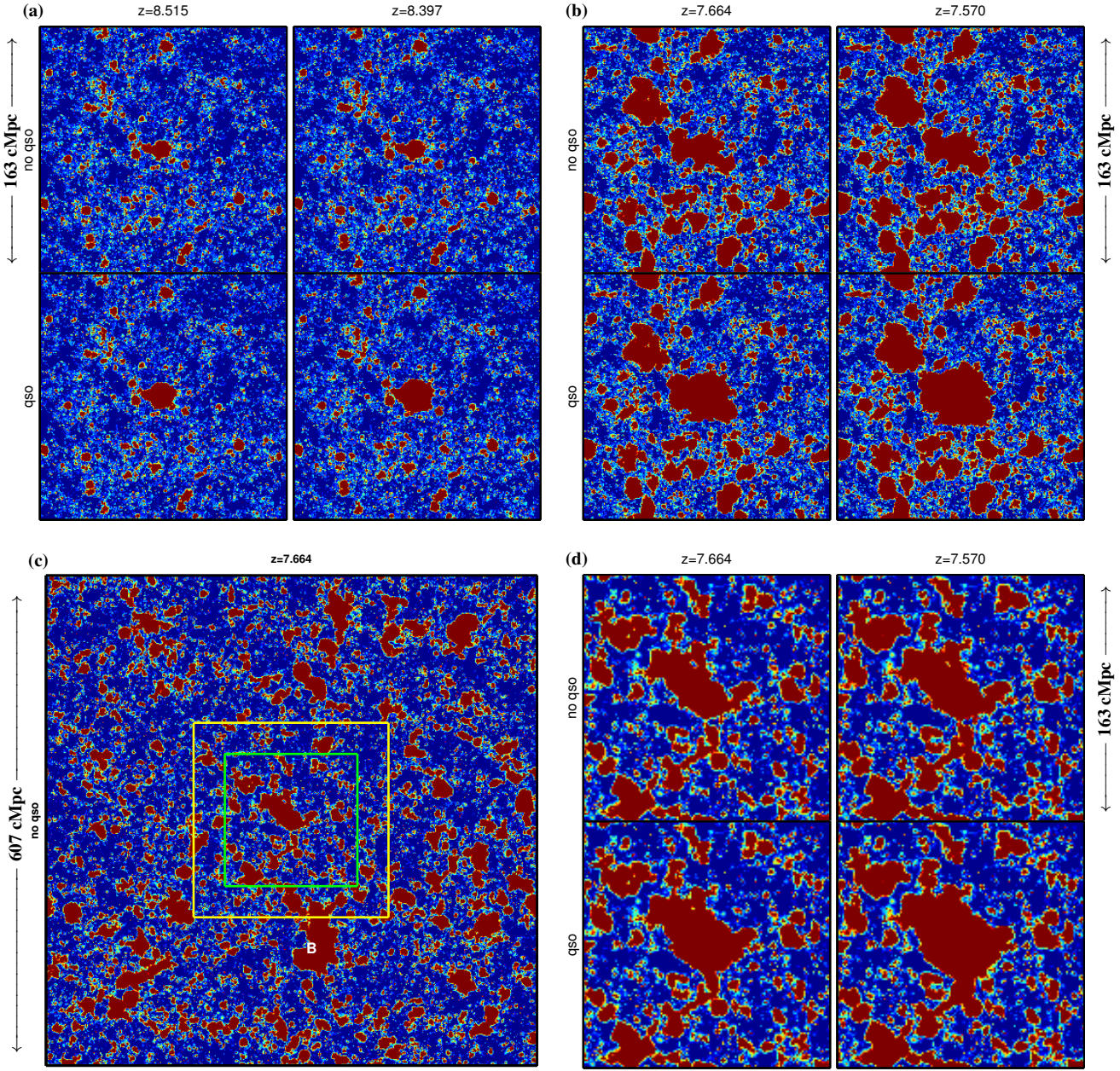

\begin{figure}
\labellist
\small\hair 2pt
\pinlabel $\textbf{(a)}$ at 15 286 
\pinlabel \rotatebox{90}{$\xleftarrow{\hspace*{1cm}}\textbf{163 cMpc} \xrightarrow{\hspace*{1cm}}$} at 455 183
\endlabellist
  \centering  
    \hspace*{-0.5cm}
\includegraphics[width=0.97\columnwidth]{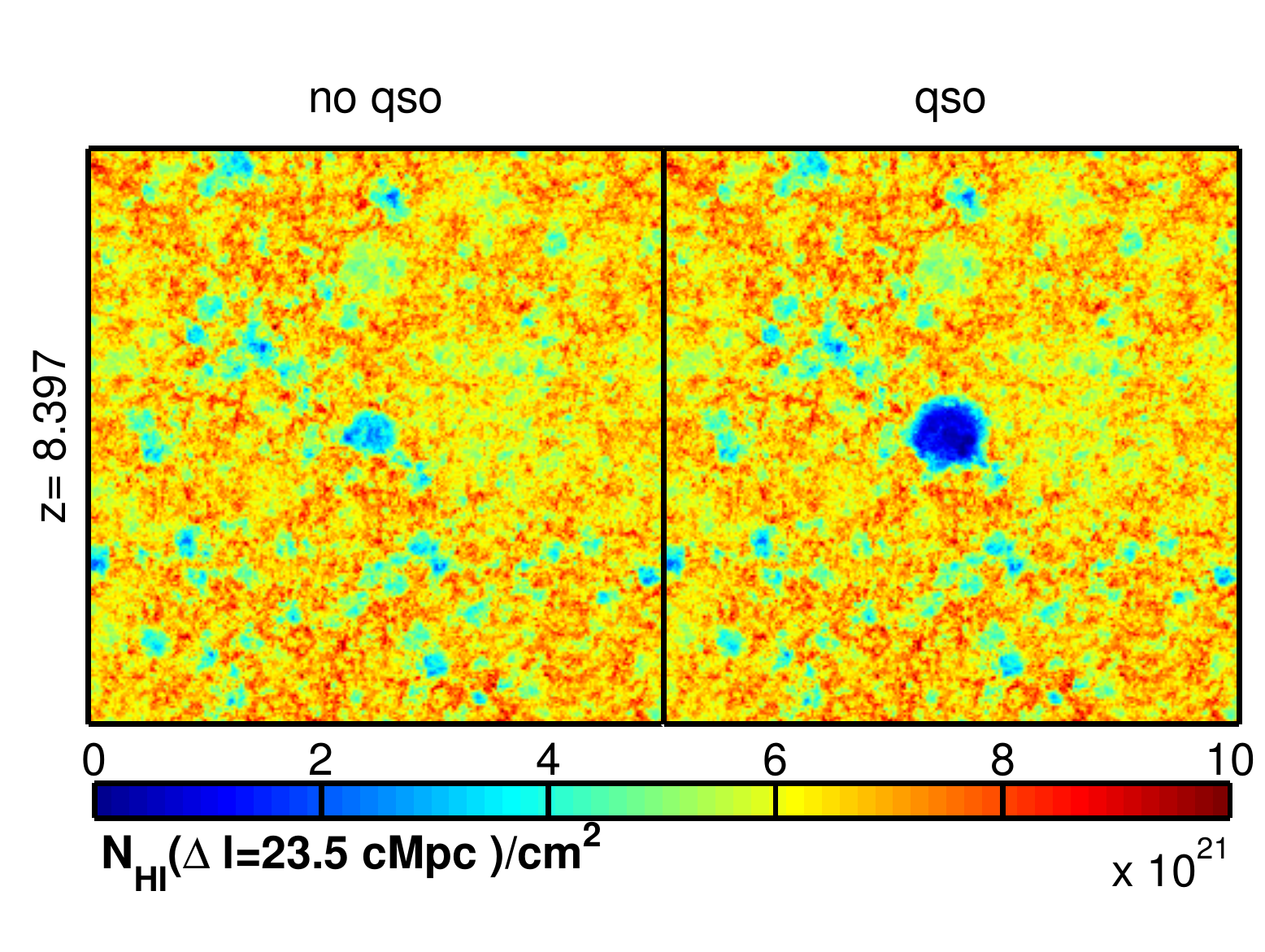} 
\labellist
\small\hair 2pt
\pinlabel $\textbf{(b)}$ at 15 286 
\pinlabel \rotatebox{90}{$\xleftarrow{\hspace*{1cm}}\textbf{163 cMpc} \xrightarrow{\hspace*{1cm}}$} at 455 183
\endlabellist
\centering
  \hspace*{-0.5cm}
\includegraphics[width=0.97\columnwidth]{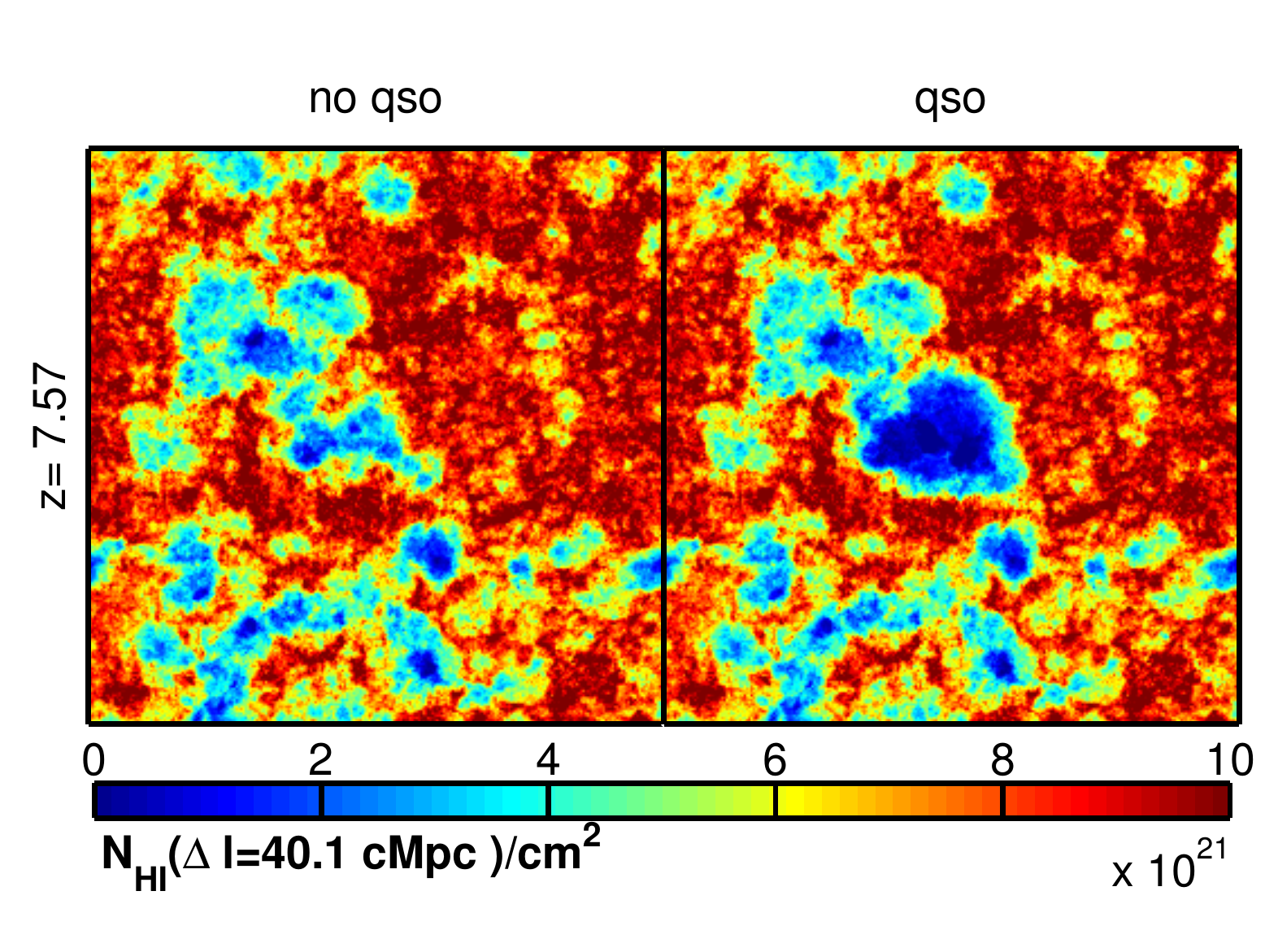}
\labellist
\small\hair 2pt
\pinlabel $\textbf{(c)}$ at 15 286 
\pinlabel \rotatebox{90}{$\xleftarrow{\hspace*{1cm}}\textbf{607 cMpc} \xrightarrow{\hspace*{1cm}}$} at 455 183
\endlabellist
  \hspace*{-0.5cm}
\includegraphics[width=0.97\columnwidth]{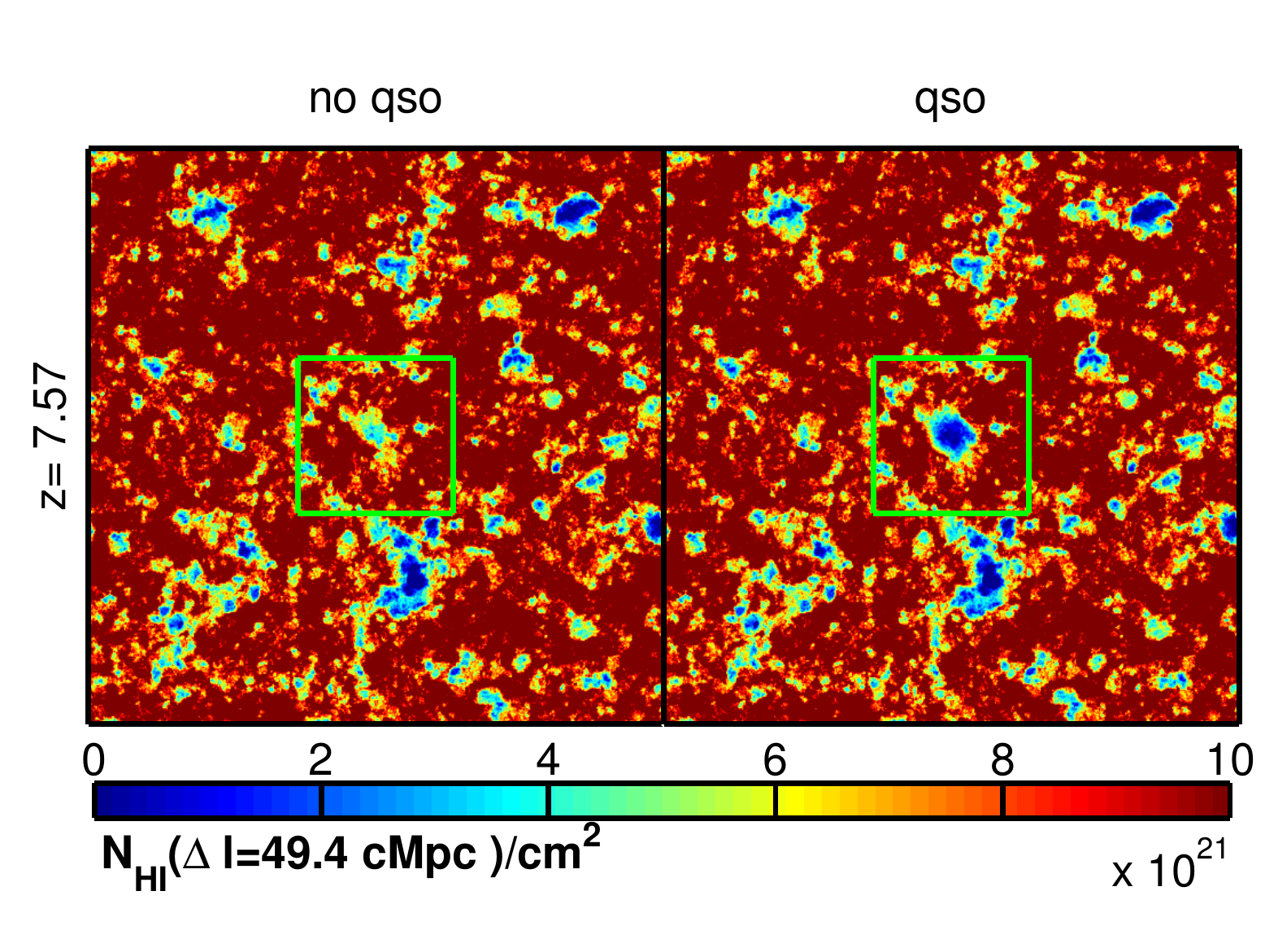}
\caption{Column densities along the line of sight over a distance roughly corresponding to the expected joint H~II region diameter as indicated in the plot. Panel (a) shows the early quasar turn on case, (b) the late quasar turn on case and (c) the large box case. The green squares in panel (c) indicate the size of the smaller simulation volume (163 cMpc).}
\label{fig:coldens}
\end{figure}
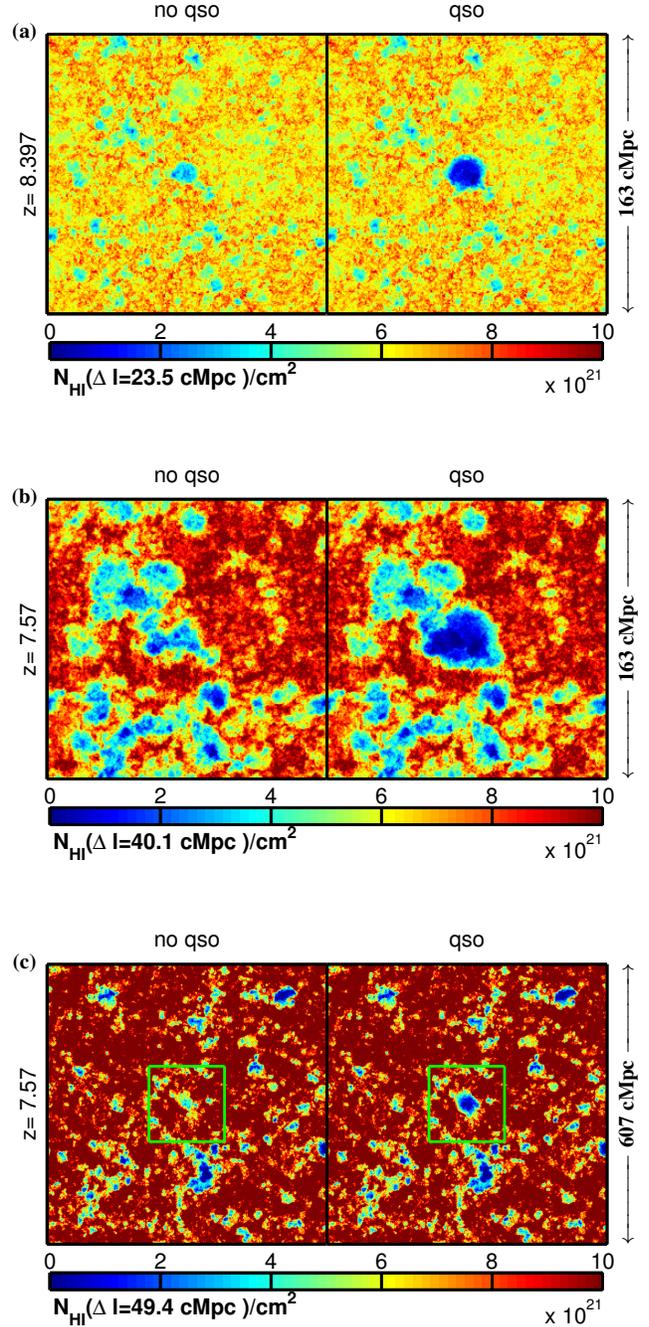

In this section, we first investigate the effect of the quasar on the size and morphology of the H~II region it is born in. We estimate the expected size of the quasar H~II region on the basis of photon counts. We compare the morphology of the H~II region around the most massive halo with and without the quasar included by means of the ionization fraction fields and column density maps. Next, we compare the quasar H~II region for one of the three cases to a large H~II region solely made by stellar sources. We show 21cm maps with a smoothing that corresponds roughly to LOFAR resolution.
To quantify the changes in size and morphology, we use a statistical measure using many lines of sight from the central source, introduced by \citet{2008MNRAS.391...63I}. Finally we analyse the He~III region around the quasar.

\subsection{Expected sizes of quasar H~II regions from photon counts}
The emitted photon rate of the quasar for each case, see Table \ref{table:summary2}, can be used to estimate very roughly the comoving radius $R$ of an expected spherical H~II region after time $t_{\rm on}$ if the surrounding medium is assumed to be completely neutral:
\bq
R=\sqrt[3]{\frac{3 t_{\rm on} \dot{N}_{\gamma}^{\rm qso}}{4 \pi n_{\rm c}}} \, ,
\label{eq:eq1}
\eq
where $t_{\rm on}$ is the time since the quasar turned on and $n_{\rm c}$ is the average comoving number density (hydrogen+helium) of the IGM. Evaluating this for the early and late quasar turn on of simulation 163Mpc\_g1.7\_8.7S and for the large box using  $t_{\rm on}=$ 23.0 Myr, gives roughly 10 cMpc, 16 cMpc and 19 cMpc, respectively.

However, all stellar sources in this region will also contribute to the ``quasar'' H~II region. For the early quasar turn on, the number of stellar photons in this region from the formation of the first source in this region to the end of the quasar phase is: 
$ N^*_{\gamma}         \sim 2.5 \times 10^{70}$. The total number of emitted ionizing photons from the quasar is 
$ N^{\rm qso}_{\gamma} \sim 2.4 \times 10^{70}$. 
In analogy, for the late quasar turn on the numbers  are  
$ N^*_{\gamma}	     \sim 9.8 \times 10^{70}$ and 
$ N^{\rm qso}_{\gamma} \sim 1.0 \times 10^{71}$. The large box simulation has 
$ N^*_{\gamma}         \sim 2.2 \times 10^{71}$ and 
$ N^{\rm qso}_{\gamma} \sim 1.7 \times 10^{71}$. These numbers are summarized in Table \ref{table:summary2}

We note that the contribution from the quasar in terms of total number of emitted photons in the region of influence of the quasar is about 50\%. We also note that the ratio $ N^*_{\gamma}/N^{\rm qso}_{\gamma}$ is largest for the most massive of our quasars. As  \citet{2008MNRAS.386.1683G} mention, the effect of suppression of low mass sources delaying reionization globally, can be reduced in highly overdense regions due to the biased formation of very massive galaxies which are not suppressed.   

Using the total number of photons, i.e. $N^*_{\gamma}+N^{\rm qso}_{\gamma}$, for estimating the comoving radius for the joined H~II region gives  $R \sim 12.0$~cMpc for the early quasar turn on and  $R \sim 19.7$~cMpc for the late quasar turn on and 25.1~cMpc for the large box. \footnote{Since this corresponds to approximately 3, 8 and 9.5 $\times 10^6$ proper lyr, respectively, the finite speed of light is not an issue since this is below the timescale we are interested in.}

\subsection{Comparisons between the  morphology of H~II regions with and without quasar}
In Fig.\ref{fig:slices}, we show for each of the three cases (early quasar turn on in panel (a), late quasar turn on in panel (b), large box in panel (d)) a slice in the plane of the most massive halo hosting the quasar at 11.5 Myr and 23.0 Myr after the quasar turn on (left and right columns of each panel, respectively). For comparison, for each case at the same times, we also show that slice in the simulation without the quasar (upper row in each panel). For better comparison, we adjusted the size of the slices from the 607Mpc\_g1.7\_8.7S simulation to a side length of 163 cMpc centred around the quasar. In panel (c), we show the full size of the (607 cMpc)$^3$ box and indicate the re-simulated region and the sub-region shown in panel (d). 

It is important to note that we show only one 2-dimensional slice through the ionization fraction field for each case. From this representation alone, no absolute judgement about the size, sphericity or connectivity of the H~II region should be made. However, we used among other things iso-surface representations and the above mentioned line-of-sight statistic (described in detail below) to confirm the statements below. It is useful to note the following from the simulations without a quasar (upper rows of panel a, b, d and panel c): 
The H~II region around the most massive halo is large compared to average H~II regions but not necessarily the largest one. Especially in the large box (panel c), there is an H~II region (marked with ``B'') which is larger. The reason for this is that there are several halos in the simulation volume which are competing for the most massive halo between $z \sim 30 \-- 6$. They are located in equally overdense regions with many clustered sources.  

From the ionization fraction fields with quasar (lower rows in panels (a), (b) and (d)), we note that the H~II regions of the quasars have as sharp boundaries as the stellar H~II regions. This is consistent with previous results by \citet{2004MNRAS.348..753S}, \citet{2008MNRAS.384.1080T} and the more detailed study of \citet{2008MNRAS.385.1561K} who found that a harder spectrum ($\alpha > -1.2$) and some level of obscuration of the low frequency quasar photons are needed to result in a detectable ionization front thickness.

Furthermore, it is visible that the growth of the H~II region during the quasar lifetime is dominated by the quasar (by construction). As we showed above, the total photon budget from the time of the formation of the first source in that region until the end of the quasar lifetime is not dominated by the quasar. We summarize the numbers is Table \ref{table:summary2}.

Next, we consider the neutral hydrogen density. We calculate the neutral column density along the line of sight for a distance corresponding to the expected H~II region diameter (from photon counts, see above) centred around the most massive halo, see Fig. \ref{fig:coldens}. 
Assuming the quasar to be observed at $z=8.397$ and $z=7.570$ respectively, the distance over which we calculated the H~I column density  corresponds to $\Delta z = 0.082, \, 0.125$  and $ 0.150$  for the early, late and large case, respectively. This corresponds to the following frequency ranges. For the early case 
(150.451 --  151.769)~MHz; for the late case (164.532 -- 166.949)~MHz and for the large box case  (164.285 -- 167.185)~MHz. This corresponds to an integration over roughly 1300, 2400 and 2900 intrinsic LOFAR frequency channels of 1 kHz. However, due to storage restrictions, LOFAR analysis will probably use channels of 100 -- 500 kHz.
 
 The size along the line of sight also implies a time difference from the far side to the quasar of about 3~Myr, 8~Myr and 9.5~Myr, for the early, late and large box case, respectively. Therefore, we actually see the far-side of the quasar H~II region at a stage that corresponds more to the one shown in the left columns of panels (a), (b) and (d) in Fig.~\ref{fig:slices}. However, we neglect this effect and consider the quasar H~II region everywhere after 23.0~Myr. The main reason for this is that our time resolution is not sufficient to make meaningful interpolations. Assuming the quasar turns off at the time of observation, the side of the H~II regions closer to us starts recombining.
However, the recombination time, even in overdense regions with $\delta \sim 10$ is still of the order of 50~Myr. For comparison, the overdensity in the H~II region is less than 0.1 in the large box case. 
We note that with our particular choice of lifetime and photon output, the part of the quasar H~II region furthest from us had already time to grow to a non-negligible size (see left columns, lower row of panels (a), (b) and (d) in  Fig.~\ref{fig:slices}). Also, less dependent on the particular quasar lifetime, the pre-existing stellar H~II region itself has a finite size.  Therefore, the anisotropy due to bubble growth might not be very important.  

In Fig. \ref{fig:coldens}, we see that the quasar H~II region in the column density map looks more distinct from other H~II regions than in the thin (0.64~cMpc for the 164~cMpc box and 1.2~cMpc for the 607~cMpc box) ionization fraction slices in Fig. \ref{fig:slices} in the sense that the difference in size is more visible. This points to a more equal extend of the H~II region in all three spatial directions. 

\begin{figure}
  \centering
\labellist
\small\hair 2pt
\pinlabel \rotatebox{90}{$\xleftarrow{\hspace*{1.0cm}}\textbf{607 cMpc} \xrightarrow{\hspace*{1.0cm}}$} at 2 188
\pinlabel \rotatebox{90}{$\textbf{qso, xy-plane}$} at 227 191
\pinlabel \rotatebox{90}{$\textbf{qso, zx-plane}$} at 2 80
\pinlabel \rotatebox{90}{$\textbf{qso, yz-plane}$} at 227 80
\endlabellist
\includegraphics[clip,width=0.97\columnwidth]{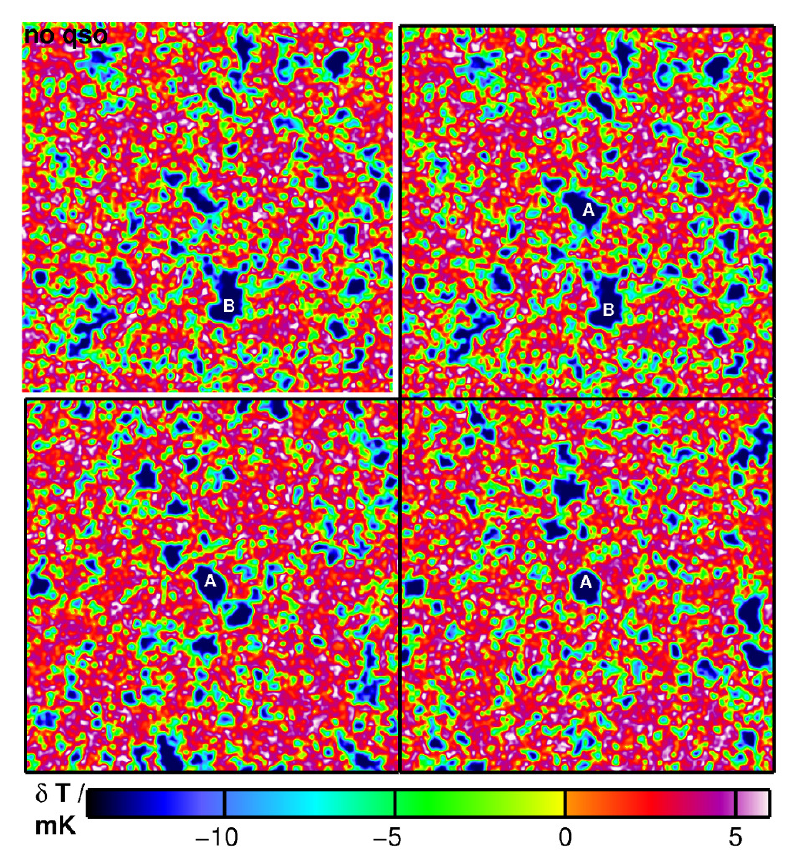}
\caption{Brightness temperature plots for the large simulation volume. Shown are three slices (xy, zx, and yz)  23.0 Mpc after the quasar turn on and one slice without the quasar included (upper left corner). The quasar H~II region is marked with ``A''. A large non-quasar region is marked with ``B'', for details see text.}
\label{fig:dT_smooth}
\end{figure}
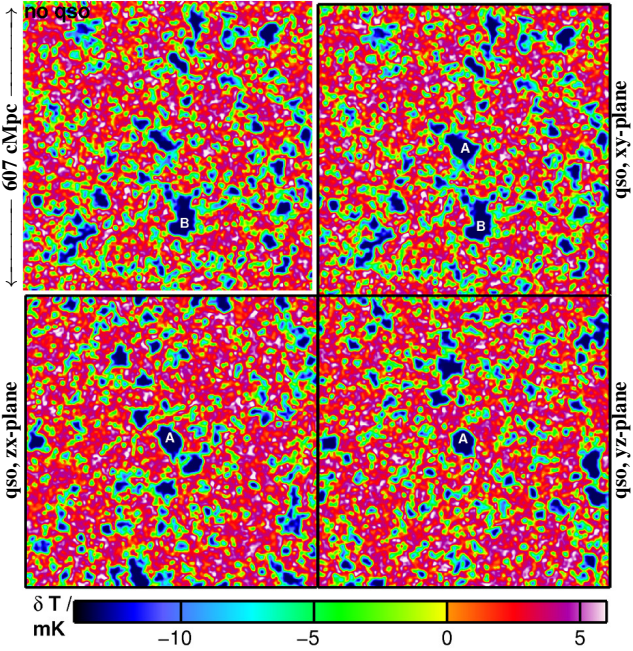

To test the 3-dimensional structure further, we show smoothed differential brightness temperature maps, see Fig.~\ref{fig:dT_smooth}. Here we concentrate on the large box case since this comprises the most luminous of our quasars and it does not by construction exclude other large scale H~II regions, see the comment in Sect.~\ref{sec:sims}. To produce those maps we assume that the IGM is heated well above the CMB temperature so that we can use Eq.~\ref{eq:dT} to convert the neutral gas density into a differential brightness temperature
\bq
\label{eq:dT}
\delta T_b= \mathcal{C}\, n_{\rm HI} (z,\rm{comoving}) \sqrt{z+1} 
\eq
with $\mathcal{C}= 4.6 \times 10^4$ K cm$^3$.
We smoothed the field with a gaussian of beam-size roughly 3 arcmin (2.75 for  $z=8.397$ and 2.5 for $z=7.57$) and subtracted the mean to mimic the effect of a missing absolute calibration typical for an interferometer. The width of the beam corresponds roughly to the LOFAR core resolution at the corresponding wavelength $\lambda\sim 1.97$~m and $\lambda \sim 1.8$~m, respectively. If LOFAR had a larger collecting area and thus a higher sensitivity, it would be possible to get maps like this.  

In Fig. \ref{fig:dT_smooth}, we show three different planes through the quasar. To assess the structure in three dimensions better, we concentrate on two sub-regions, the quasar region \textbf{A} and the large H~II region marked as \textbf{B}. Region \textbf{B} is not a quasar H~II region but formed in a highly biased region with many massive halos. Since there is no central source that dominates the H~II region, we first find the centre cell of this region by minimizing the sum of brightness temperatures in a sphere with the same radius as the quasar H~II region.  

\begin{figure}
  \centering
\labellist
\small\hair 2pt
\pinlabel $\textbf{(a)}$ at -5 221 
\pinlabel \rotatebox{90}{$\xleftarrow{\hspace*{0.4cm}}\textbf{43.9 cMpc} \xrightarrow{\hspace*{0.4cm}}$} at 448 142
\endlabellist
\includegraphics[clip,width=\columnwidth]{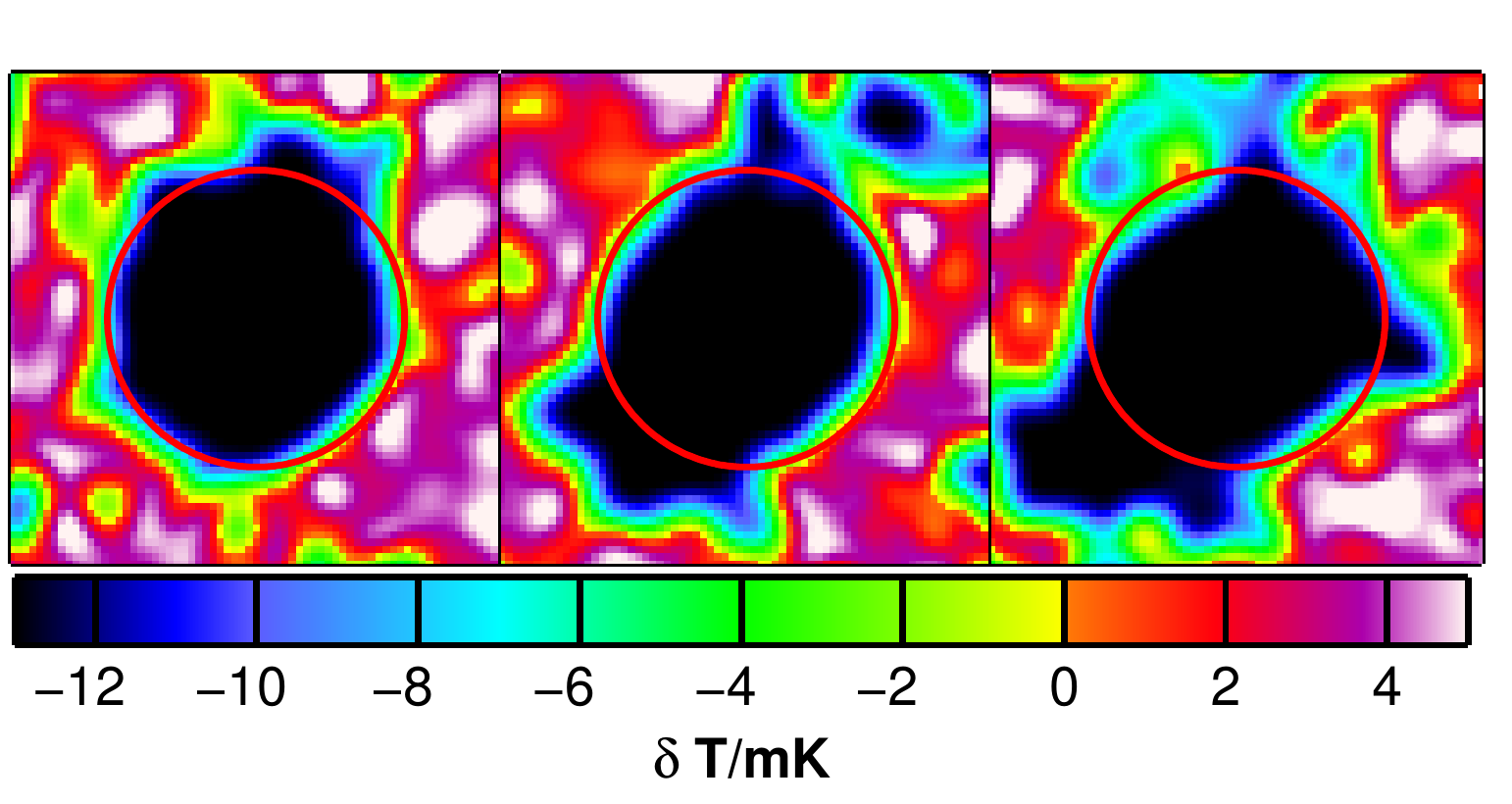} \\
\labellist
\small\hair 2pt
\pinlabel $\textbf{(b)}$ at -5 221
\pinlabel \rotatebox{90}{$\xleftarrow{\hspace*{0.4cm}}\textbf{43.9 cMpc} \xrightarrow{\hspace*{0.4cm}}$} at 448 142
\endlabellist
\includegraphics[clip,width=\columnwidth]{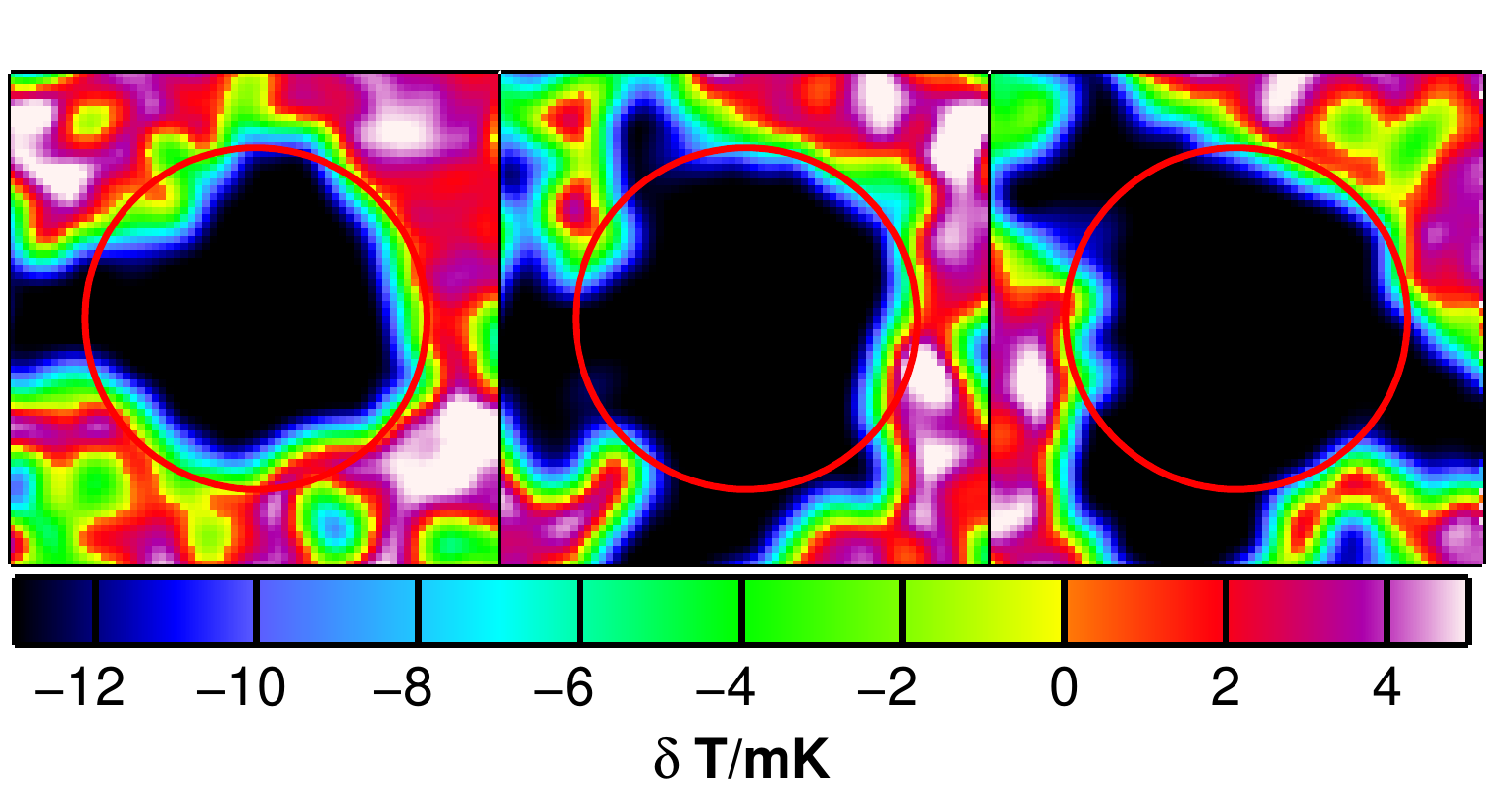}
\caption{Brightness temperature maps as in Fig. \ref{fig:dT_smooth}, colour coded as indicated by the colour bar. Shown are three slices perpendicular to the principle axes for each H~II region. Panel (a) shows the quasar H~II region, region \textbf{A}, panel (b) shows region \textbf{B}. The side length of each slice is 43.9 cMpc. The circles indicate the size of the H~II region found by the matched filter method, see Sect. \ref{sec:results2}. Note that region \textbf{B} is not in the same planes as shown in Fig. \ref{fig:dT_smooth}}
\label{fig:dT_3D}
\end{figure}

\begin{figure}
  \centering
\includegraphics[clip,width=\columnwidth]{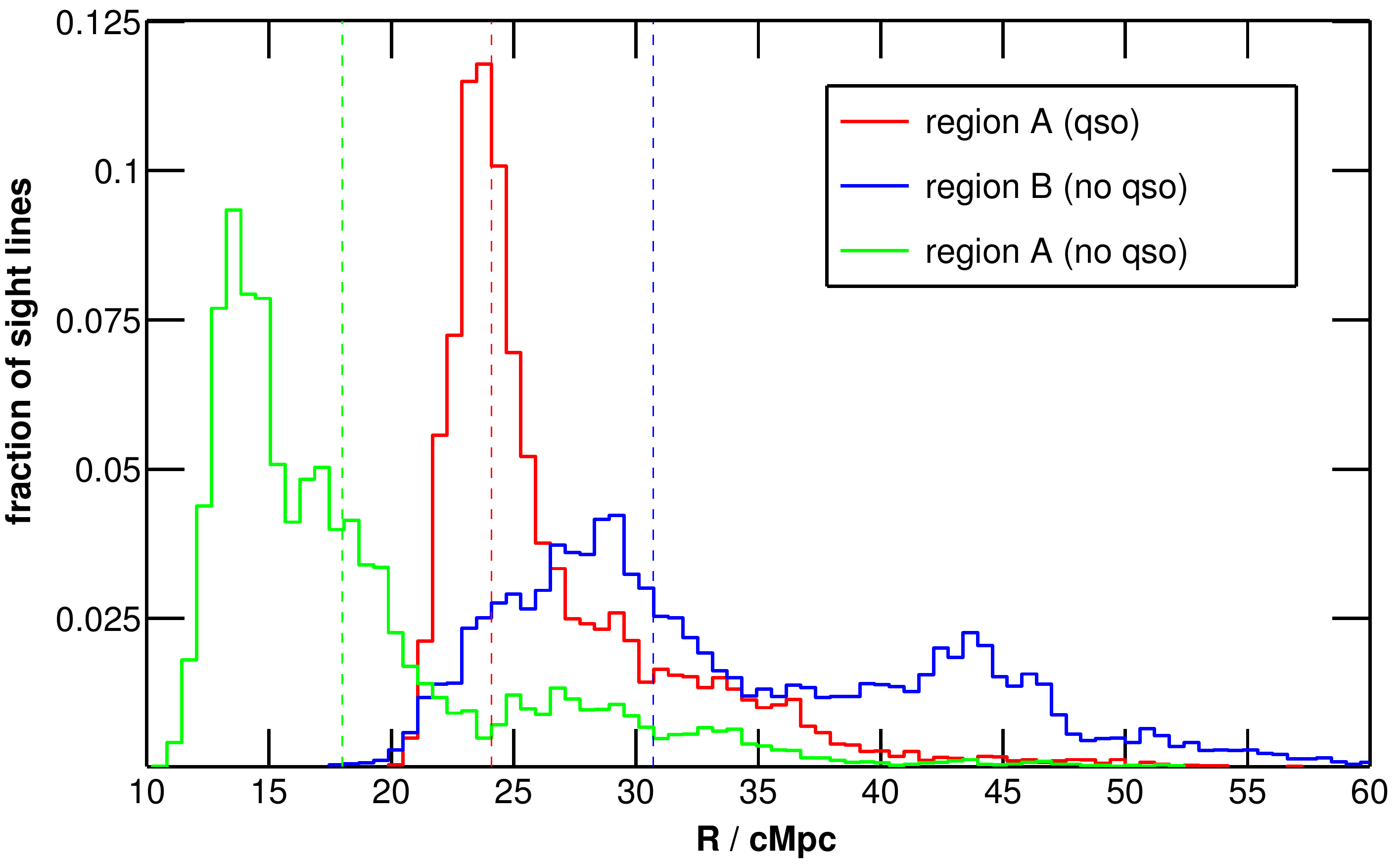} 
\caption{Histogram of H~II region sizes found by casting rays from the centre and registering where the optical depth along the ray surpasses $\tau_{\rm s}$. Shown are results for region \textbf{A} with (red) and without (green) the quasar, and region \textbf{B} (blue). The threshold optical depth for all cases is $\tau_s=4.6$. The dashed lines indicate the medians.}
\label{fig:bubble_stat}
\end{figure}

In Fig. \ref{fig:dT_3D}, we show slices perpendicular to the three principal axes for both regions indicated in Fig.~\ref{fig:dT_smooth}. Panel (a) shows the quasar region \textbf{A} and panel (b) shows region \textbf{B}. The circles have  the radii found for the respective regions with the matched filter method, see Sect.~\ref{sec:results2}. It is visible that the radius found by the matched filter method allows for some emission inside the bubble. The size found for region \textbf{B} is larger than for region \textbf{A}, $R_{\rm b}(B)\sim 28.7$~cMpc compared to $R_{\rm b}(A)\sim 24.9$~cMpc. However, as will be discussed in Sect. \ref{sec:results2}, the peak in the signal-to-noise ratio is not as distinct, see panel (c) in Fig.~\ref{fig:snr-noqso1}, which is perhaps obvious given the less spherical shape of region \textbf{B} as visible in Fig.~\ref{fig:dT_3D}. To quantify the degree of sphericity, we use a method introduced by \citet{2008MNRAS.391...63I}. Here, rays are cast from the central source and for each ray, the distance where the optical depth $\tau$ along this ray surpasses a certain limit $\tau_{\rm s}$ is recorded. In Fig. \ref{fig:bubble_stat}, we show histograms for these distances for region \textbf{A} with and without a quasar, as well as for region \textbf{B}. As the central point for region B, we chose the same point as for the matched filter analysis. For each point, we cast $10\,000$ rays and use the threshold optical depth $\tau_{\rm s}=4.6$. We also tested the smaller threshold $\tau_{\rm s}=1$. We found the curves to be very similar, confirming the visual impression that the edges of the H~II regions are sharp. Comparing the H~II region around the most massive halo (\textbf{A}) with and without quasar, we see that that the quasar H~II regions is a bit more spherical and larger in size. The non quasar H~II region \textbf{B} is larger than the quasar H~II region but much less spherical.

\begin{figure}
  \centering
  \labellist
\small\hair 2pt
\pinlabel \rotatebox{90}{$\xleftarrow{\hspace*{2.2cm}}\textbf{163 cMpc} \xrightarrow{\hspace*{2.2cm}}$} at 0 164
\endlabellist
\includegraphics[clip,width=\columnwidth]{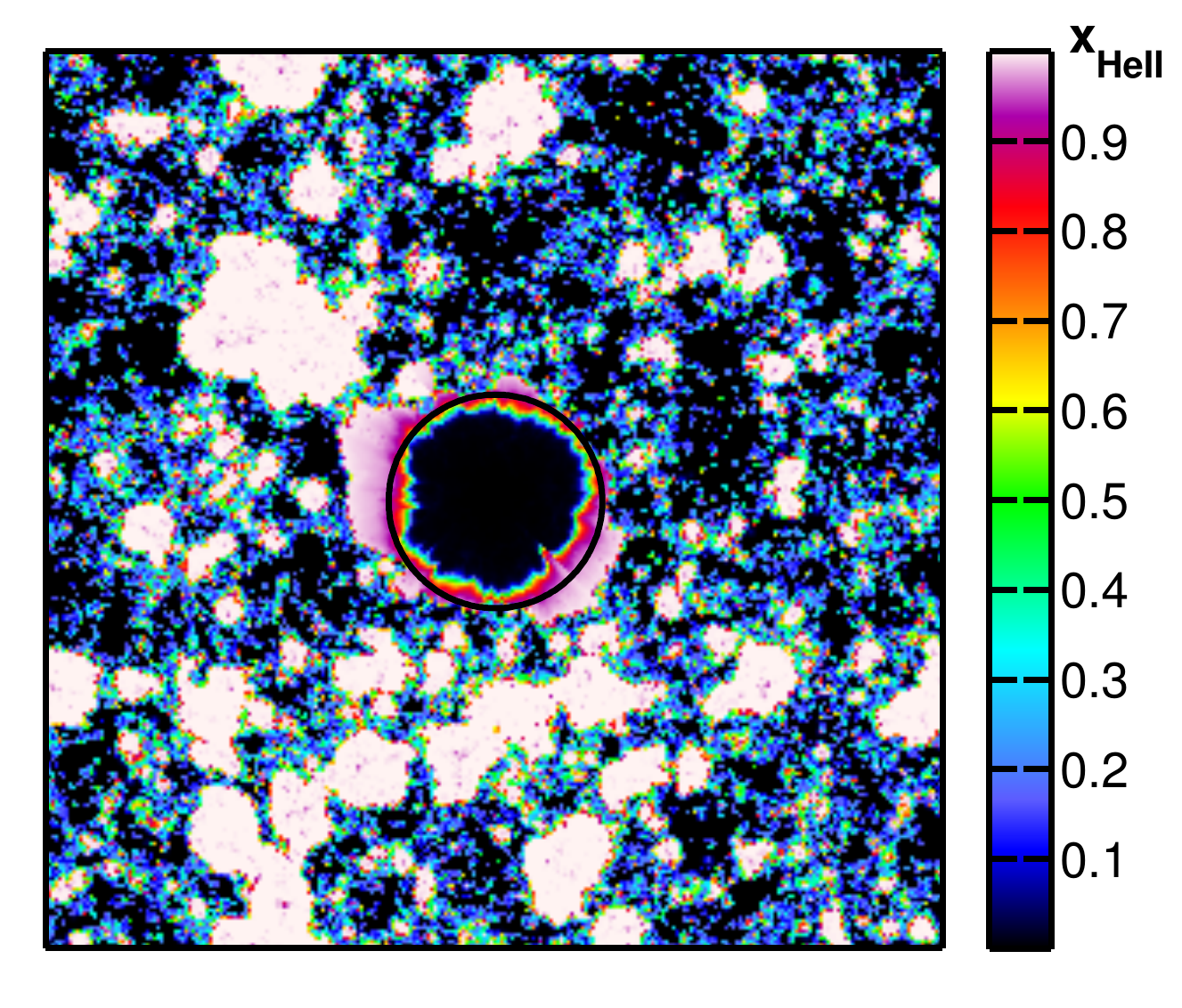} 
\caption{He~II fraction for the late quasar turn on case, colour coded as indicated by the colour scale. The circle indicates the radius calculated on the basis of photon counts, see text.}
\label{fig:HeII}
\end{figure}

\subsection{Helium ionization region around the quasar}
In Fig. \ref{fig:HeII} we show the He~II fraction for the late quasar turn on. We chose this case and not the large box, since it has a higher resolution. The circle indicates the calculated He~III radius of the quasar, $R\sim 19.1$ cMpc, assuming all photons able to double ionize helium ($N_{\gamma, \, >54.4 \, \rm{eV}}^{\rm qso}=1.3 \times 10^{70} $) do so. Since the fraction of photons going into the ionization of a species depends on its contribution to the optical depth, this is an overestimation. 

There are mainly two differences between the H~II region and the He~III region: \\
(1) Since the contribution to photons with energies above 54.4 eV (the threshold energy of helium double ionization) of stellar sources is negligible, the quasar dominates the double-helium ionizing photon output. There are no other large-scale He~III regions to merge with. Therefore, the He~III region is much more spherical than the H~II region. \\
(2) The He~II \-- He~III front is much shallower than the H~II region front. 

As discussed by \citet{2009PhRvD..80f3010M} and \citet{2009arXiv0905.1698B}, the hyperfine transition line of $^3{\rm He ~II}$ at 8.7 GHz could be in principle used to detect singly ionized helium. However, as \citet{2009PhRvD..80f3010M} calculate, the spin temperature of the transition is not significantly decoupled from the CMB temperature. Therefore, this line will only be observable in absorption against, for example, another background quasar \citep{2009arXiv0905.1698B}.

\section{Simulating Visibilities}
\label{sec:visibilities}
\begin{figure}
\includegraphics[width=\columnwidth, angle=0]{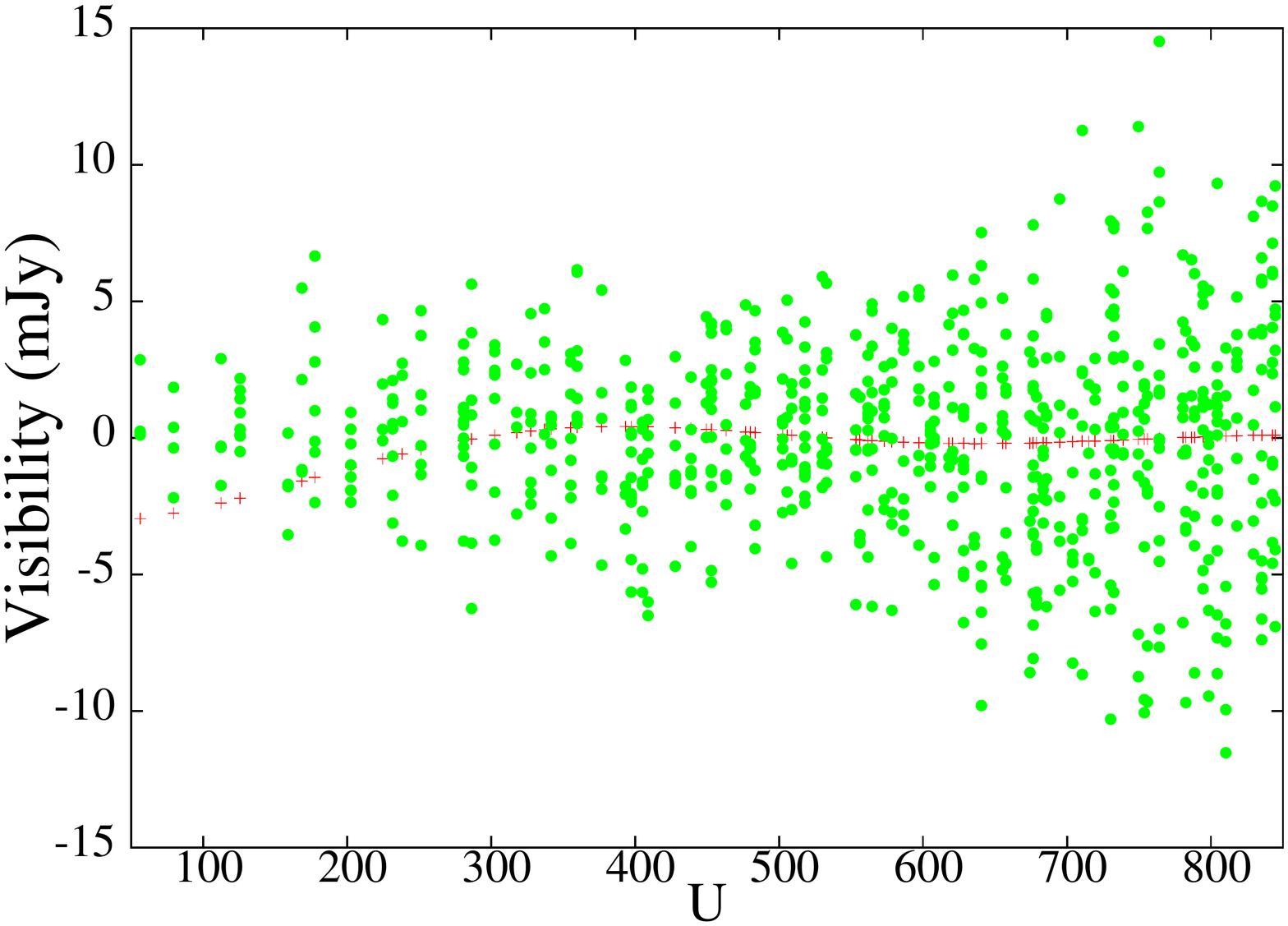}
\caption{Comparing of the signal in the visibility from a single quasar H II region and  $1200$ hours LOFAR system noise.  Green circles represent the noise in the real part of the visibility as a function of baseline $U$ for $1200$ hours of observations in a single frequency channel of width $38 \, {\rm KHz}$ at redshift 7.57. Red crosses are 20 times the expected signal from a single H II region of radius $20\, {\rm cMpc}$ embedded in a uniform H I density field with $x_{\rm HI}=0.55$ at redshift 7.57. }
\label{fig:sigandnoise}
\end{figure}

The quantity measured in radio-interferometric observations is the
visibility $V(\u, \nu)$ which is related to the specific intensity pattern
on the sky $I_{\nu}(\th)$  as
\be
V(\u, \nu)=\int d^2 \theta A(\th) I_{\nu}(\th) e^{2 \pi i \th . \u}
\label{eq:vis-tot}
\e
Here the baseline $\u = \vec{d}/\lambda$ denotes the antenna separation $\vec{d}$ projected in the plane perpendicular to the line of sight in units of
the observing wavelength $\lambda$ corresponding to the observing frequency $\nu$, $\th$ is a two dimensional vector in the
plane of the sky with origin at the centre of the field of view (FoV), and $A(\th)$
is the beam pattern of the individual antenna. For LOFAR, we approximate $A(\th)$ by a top-hat function with FoV $5 \degr$.

The visibility recorded in radio-interferometric observations of a H II region during the EoR is actually a combination of several contributions
\be
V (\u , \nu) = S(\u , \nu) + HF (\u, \nu) + F (\u , \nu) + N (\u , \nu),
\label{eq:vis-tot1}
\e
where $S(\u , \nu)$ is the signal from an H II region of comoving radius $R_b$ embedded in an uniform H I distribution. ${ HF} (\u, \nu)$ is the contribution from the fluctuating H I outside the H II region which arise because of the existence of other H II regions and density fluctuations in the H I. $F (\u , \nu)$ and $N (\u , \nu)$ are the contributions from the foreground and the system noise, respectively. The H I fluctuations cannot be reduced by increasing the observing time and thus put a fundamental lower limit on the size of H II regions that can be detected \citep{2007MNRAS.382..809D, 2008MNRAS.391.1900D}. The contribution from the fluctuating H I is less than the signal if the targeted H II region is bigger than the average size of surrounding H II regions \citep{2008MNRAS.391.1900D}. Foregrounds $F (\u , \nu)$ are much stronger (several orders of magnitude) than the signal, but they are expected to be smooth in frequency. In Sect. \ref{sec:foregrounds} we show that foregrounds can be subtracted out and the residual foreground is below the signal for  LOFAR \citep[also see][ for the GMRT and MWA]{2008MNRAS.391.1900D}. In the next section we discuss the signal in visibility and how we calculate it from our simulated image cubes.

\begin{figure*}
\centering
\labellist
\small\hair 2pt
\pinlabel $\rotatebox{90}{\textbf{(a)}}$ at -2 15 
\endlabellist
\rotatebox{270}{\includegraphics[height=.45\textwidth]{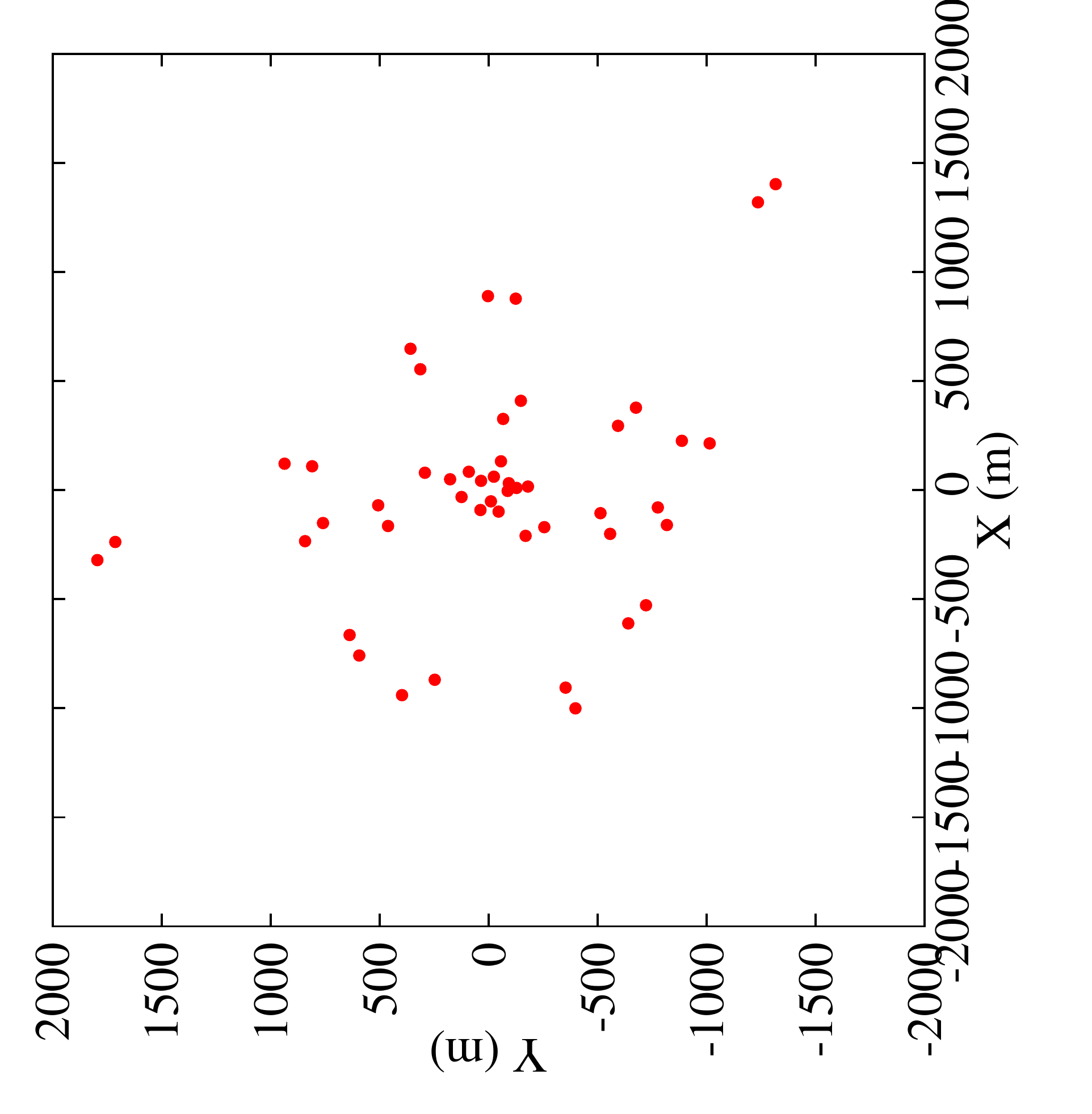}} \hspace{12mm}
\labellist
\small\hair 2pt
\pinlabel $\rotatebox{90}{\textbf{(b)}}$ at -2 0 
\endlabellist
\rotatebox{270}{\includegraphics[height=0.45\textwidth]{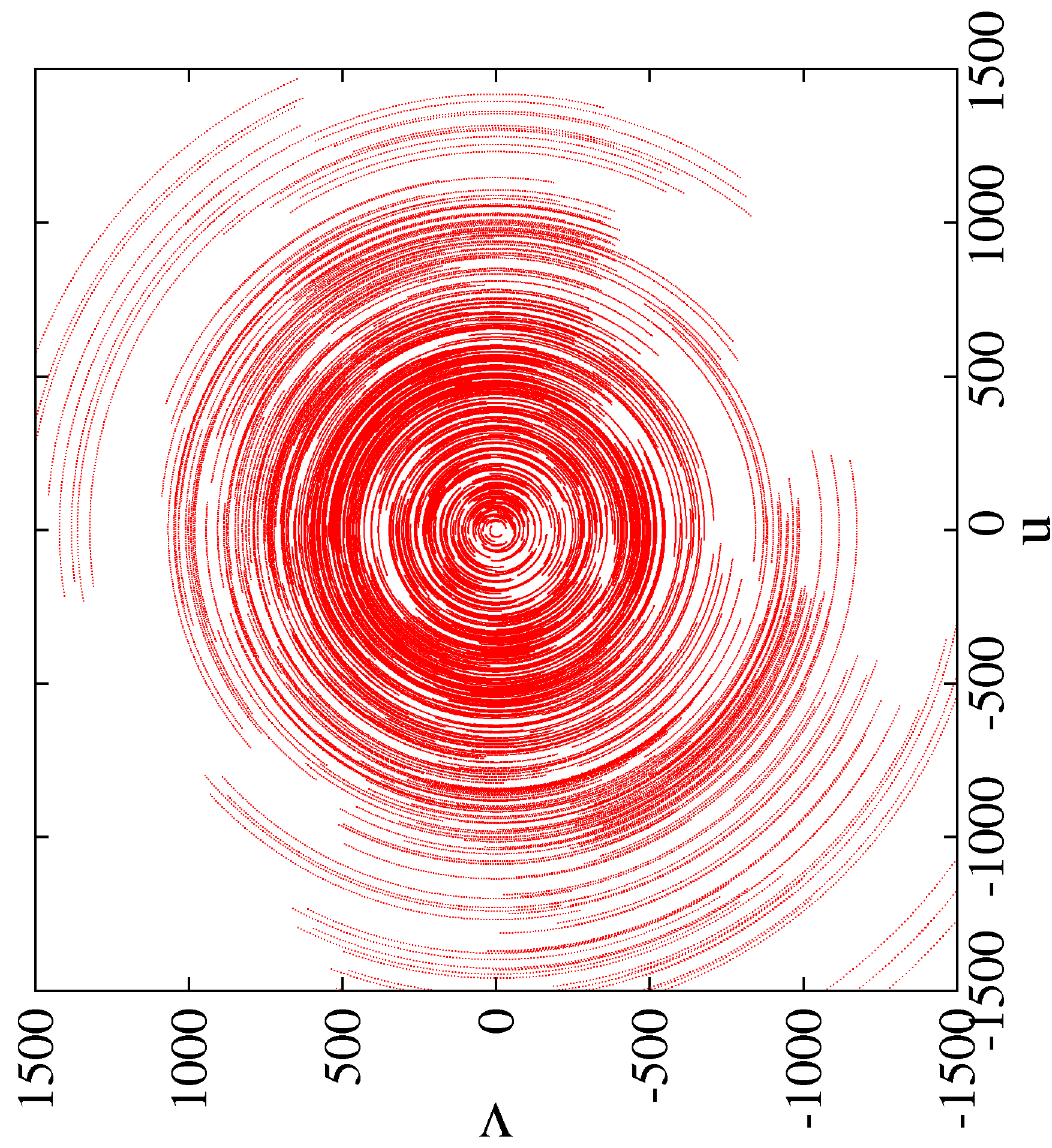}}
\caption{LOFAR coordinates of the $48$ core stations (panel a). The LOFAR uv tracks for a $4$ hours observations towards the NCP (panel b).}
\label{fig:coordinate}
\end{figure*}

\subsection{Simulating Signal}
The expected visibility pattern for a simplistic case such as a spherical H II region embedded in an uniform H I density can be calculated analytically \citep{2007MNRAS.382..809D}. A single frequency channel will observe a circular ionized region across the H II region. The visibility for any circular ionized region of radius $R_{\nu}$ at a comoving distance $r_{\nu}$ can be written as,
\be \resizebox{.9\columnwidth}{!}{$
S(\u, \nu)=- \pi \bar{I}_{\nu} x_{\rm HI} \theta^2_{\nu} e^{2 \pi i \u.\th_{\rm c}}\left[ \frac{2 J_1(2 \pi U \theta_{\nu})}{2 \pi U \theta_{\nu}}\right] \Theta \left ( 1-\frac{|\nu-\nu_b|}{\Delta \nu_b}\right) $}
\e
where $\bar{I}_{\nu}$ is the mean specific intensity for the mass average neutral fraction  $x_{\rm HI}=1$, $\theta_{\nu}=R_{\nu}/r_{\nu}$ is the angular size of the circular ionized region across the H II region. The phase factor $e^{2 \pi i \u.\th_{\rm c}}$ arises because of arbitrary  position of H II region $\th_{\rm c}$ with respect to the centre of the FoV. $J_1 (x)$ and $\Theta(x)$ are the first order Bessel function and Heaviside step function, respectively. The radius $R_{\nu}$ of any circular region through the H II region is calculated from the H II region radius $R_{\rm b}$ by $R_{\nu}=R_{\rm b} \sqrt{1-(\Delta \nu/\Delta \nu_{\rm b})^2}$ where $\Delta \nu=\nu-\nu_{\rm c}$ is the distance of the circular region from the H II region centre $\nu_{\rm c}$ and $\Delta \nu_{\rm b}=R_{\rm b}/\frac{dr_{\nu}}{d\nu}$ is the H II region size in frequency units. The U range, frequency range and peak value of a signal $S(\u, \nu)$ scale as ${R_{\rm b}}^{-1} $, $R_{\rm b}$ and $R_{\rm b}^2$, respectively i.e, for a larger H II region the signal is confined to shorter baselines. Fig. \ref{fig:sigandnoise} plots  the expected signal $S(\u, \nu)$ multiplied by $20$ as a function of baseline $U$ for the central circular ionized map of a single H II region of radius $20\,{\rm cMpc}$ embedded in a uniform H I density with $x_{\rm HI}=0.55$ at redshift $7.57$. For further details we refer the reader to  figures  1 \& 2 and the discussion in the section 2.1 in \citet{2007MNRAS.382..809D}. In our current simulations described in Sect. \ref{sec:results1} we see that the quasar H II region is not spherical and the outside H I is fluctuating. However the discussion above is still useful to understand how the signal looks as a function of baseline $U$ and frequency $\nu$. In addition, the filter we consider later is based on the signal we discussed above. 

The radiative transfer simulations provide us with the brightness temperature distribution $ \delta T_{\rm b} (\vec {\theta}, \nu)$ at different angular locations $\vec {\theta}$ and frequencies $\nu$ assuming a global heating as described in Sect. \ref{sec:results1}. The `early' and `late' simulation cubes which have a side length $163 \, {\rm cMpc}$ are $ \sim 1^{\circ}$ and $\sim 10$  MHz wide in angle and frequency, respectively. For the large box simulation, these numbers are $ \sim 4^{\circ}$  and $\sim 36 \, {\rm MHz}$. The image $\delta T_{\rm b} (\th, \nu)$ at each frequency channel is multiplied by $\frac{d B_{\nu}}{dT}=\frac{2 k_{\rm B} \nu^2}{c^2}$ to convert to the specific intensity $I_{\nu}(\th)$. Next we perform the 2D discrete Fourier transform (DFT) of the product to simulate the complex visibilities. The LOFAR core has baselines in the range $20 < U < 2000$ which is adequate to capture the H I signal from H II regions which are expected to be confined to small baselines $U < 1000$ for the stages of reionization we consider here.

\subsection{Simulating Noise}
\label{sec:noise_sim}
In this section we describe our method of simulating the expected noise for LOFAR in the visibility. 

The system noise contribution $N (\u ,\nu)$ (see Eq. \ref{eq:vis-tot1}) in each baseline and frequency channel is expected to be an independent Gaussian random variable with zero mean ($\langle \hat{N}\rangle = 0$) and whose rms is independent of $\u$ . The predicted rms noise contribution in the real (or imaginary) part  of a single visibility for a single polarization can be written as  \citep{2001isra.book.....T}
\be
N^{\rm R}_{\rm rms}=\frac{\sqrt{2}  \eta k_{\rm B} T_{\rm sys}}{ A_{\rm eff}\sqrt{\Delta \nu \Delta t}}
\label{eq:noise1}
\e
where $T_{\rm sys}$ is the total system temperature, $k_{\rm B}$ is the Boltzmann constant and $\eta$ is the antenna efficiency. We assume $\eta$ to be 1. $A_{\rm eff}$ is the effective collecting area of each antenna, $\Delta \nu$ is the frequency channel width and $\Delta t$ is the correlator integration time. Eq.
\ref{eq:noise1} can be rewritten as
\be \resizebox{.98\columnwidth}{!}{$
N^{\rm R}_{\rm rms} =1.95 \,\rm{ Jy} \left(\frac{T_{\rm sys}}{500 K}\right) \left(\frac{A_{\rm eff}}{500 \rm{m}^2}\right)^{-1} \left (\frac{\Delta \nu}{\rm{MHz}}\right)^{-1/2}\left (\frac{\Delta t}{\rm{s}}\right )^{-1/2} $}
\e
For  LOFAR we assume 
\bqs
T_{sys}=T_{\rm sky}+T_{\rm receiver}=[60\,(\nu/300 MHz)^{-2.55}+140]\, \rm{K}
\eqs
 \citep{2009MNRAS.397.1138H}
 and 
 $A_{\rm eff}=526\,(150 \, \rm{MHz}/ \nu)^2 \, \rm{m}^2$. \footnote{http://astron.nl/radio-observatory/astronomers/lofar-imaging-capabilities-sensitivity/sensitivity-lofar-array/sensiti}  For two polarizations the noise is reduced by a factor of $\sqrt{2}$. Our small and large simulated cubes have resolutions $0.64 \,{\rm cMpc}$ and $1.2 \, {\rm cMpc}$ which are equivalent to frequency resolutions $\Delta \nu \sim 38\, {\rm KHz}$  and $\sim 71 \,{\rm KHz}$  respectively at redshift $z=7.57$. For $\Delta \nu =70 \, {\rm KHz}$, integration time $\Delta t=100$~s, this gives $N^{\rm R}_{\rm rms}=608 \,$ mJy for two polarizations at redshift $z=7.57$. 

The baselines obtained using DFT of our simulations are uniformly distributed on a $uv$ plane. In real observations, the baselines will have a complicated distribution depending on the antenna layout and the direction of observations. We use 48 high band antennae (left panel of Fig. \ref{fig:coordinate}) of the LOFAR core to simulate the uv tracks for $4$ hours of observations with $100$s integration time at declination $90^{\circ}$ (NCP\footnote{One of the potential fields for the LOFAR -EoR }) (see right panel of Fig. \ref{fig:coordinate}). We then count the total number of measurements $n(\u)$ in the area defined by $u \pm \Delta u/2$ and $v \pm \Delta v/2$ around the baseline $\u (u,v)$ obtained using DFT of our simulations, where $\Delta u \Delta v$ is the grid size determined by the simulated cube size. We then reduce the noise rms in both the real and imaginary part to $N^{\rm R}_{\rm rms}/\sqrt{n(\u)}$.  We generate a Gaussian random number with rms $N^{\rm R}_{\rm rms}/\sqrt{n(\u)}$ for each grid point. We assume the signal and noise to be zero where there are no measurements. We do not expect the rms  to change much for different choices of integration time. Since we show our results for a total observing time of $1200$ hours, we further reduce the noise by a factor of $\sqrt{300}$.  

Finally we simulate $1000$ independent noise realizations to investigate how the system noise changes the H II region size determination and quantify statistically the accuracy at which different H II region sizes can be measured. We discuss the effect of system noise on the size determination of a quasar H II region in Sect. \ref{sec:noise}.

\section{Matched filter formalism: A brief summary}
\label{sec:matched_filter}
The details of the matched filter technique were discussed in  \citet{2007MNRAS.382..809D, 2008MNRAS.391.1900D}. Here we briefly describe the main features relevant for our study. The signal component $S(\u , \nu)$ in the observed visibilities $V (\u, \nu)$ is expected to be buried deep in other contributions (see Eq. \ref{eq:vis-tot1}), some of which are orders of magnitude larger. In order to enhance the signal to noise ratio (SNR) we sum the entire observed visibility signal weighed with the filter. The estimator $\hat{E}$ is defined as,
\be
\hat{E}=\left [ \sum_{\rm{a,b}}S_{\rm f}^*(\u_{ a},\nu_{b})\hat{V}(\u_{ a},\nu_{b}) \right]/\sum_{ a,b}
\e
where $S_{\rm f} (\u , \nu)$ is a complex filter which has been constructed to detect a particular H II region and the `*' denotes the complex conjugate, $\hat{V} (\u_{ a} , \nu_{b} )$ refers to the observed visibilities and $\u_{ a}$ and $\nu_{b}$ refer to the different baselines and frequency channels in the observation. The filter $S_{\rm f} (\u , \nu)$ is a function of the comoving radius $R_{\rm f}$, redshift $z_{\rm f}$ and angular position  $\th_{\rm f}$. We do not show this explicitly, the values of these parameters will be clear from the context.
 
Using the simulated visibilities, we evaluate the estimator as
\be
\hat{E}=(\Delta u \Delta v \Delta \nu)\sum_{i,j}S_{\rm f}^*(\u_i,\nu_j)\hat{V}(\u_i,\nu_j) \rho_N(\u_i,\nu_j)
\e
where the sum is now over the baselines and frequency channels in the simulation. $\Delta u $, $\Delta v $, $\Delta \nu$ are the grid sizes along u,v and $\nu$ axis obtained from the simulation. The product $\Delta u \Delta v \Delta \nu \rho_N(\u ,\nu)$ is the fraction of all baselines in the interval $\vec{U}$, $\vec{U}+\vec{dU}$ and $\nu$, $\nu+d\nu$. Note that the  normalized baseline distribution function $\rho_N(\u , \nu)$ is usually frequency dependent, and it is normalized so that $\int d^2 U d\nu \rho_N(\u , \nu) = 1$. 

\begin{figure*}
\labellist
\small\hair 2pt
\pinlabel \rotatebox{90}{$\textbf{(a)}$} at 40 655
\endlabellist
\rotatebox{270}{\includegraphics[height=.33\textwidth]{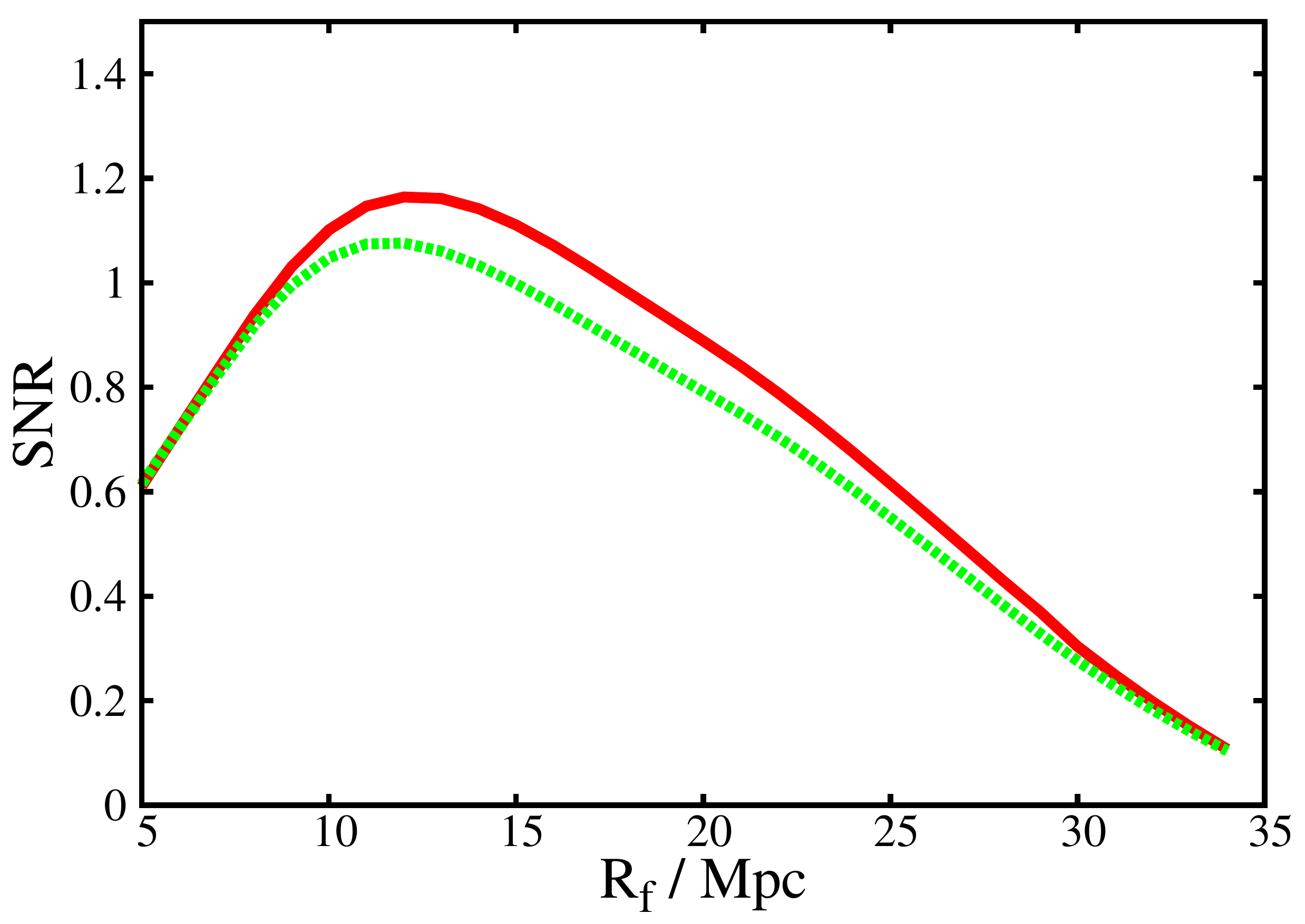}}
\labellist
\small\hair 2pt
\pinlabel \rotatebox{90}{$\textbf{(b)}$} at 40 655
\endlabellist
\rotatebox{270}{\includegraphics[height=.33\textwidth]{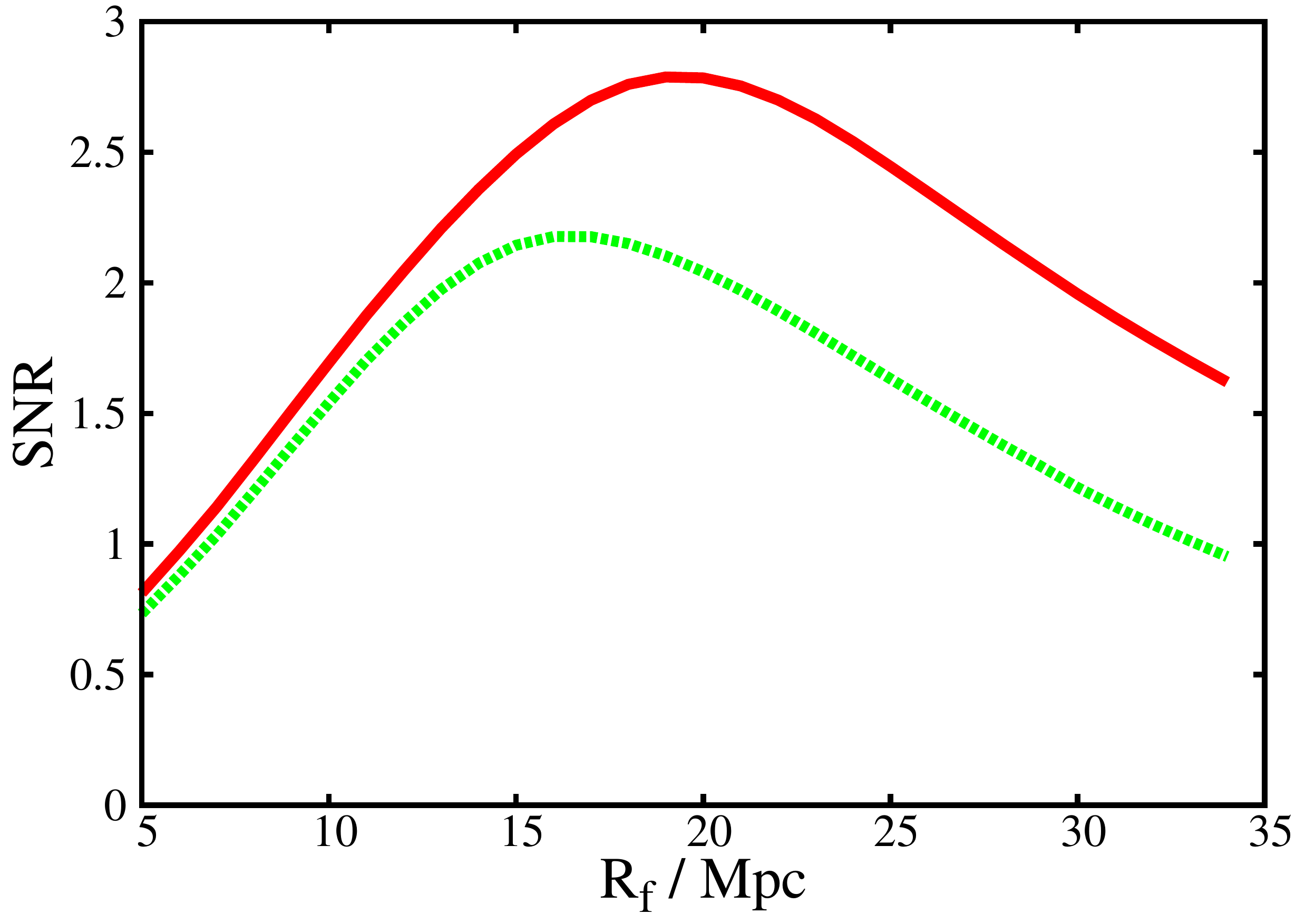}}
\labellist
\small\hair 2pt
\pinlabel \rotatebox{90}{$\textbf{(c)}$} at 40 655
\endlabellist
\rotatebox{270}{\includegraphics[height=.33\textwidth]{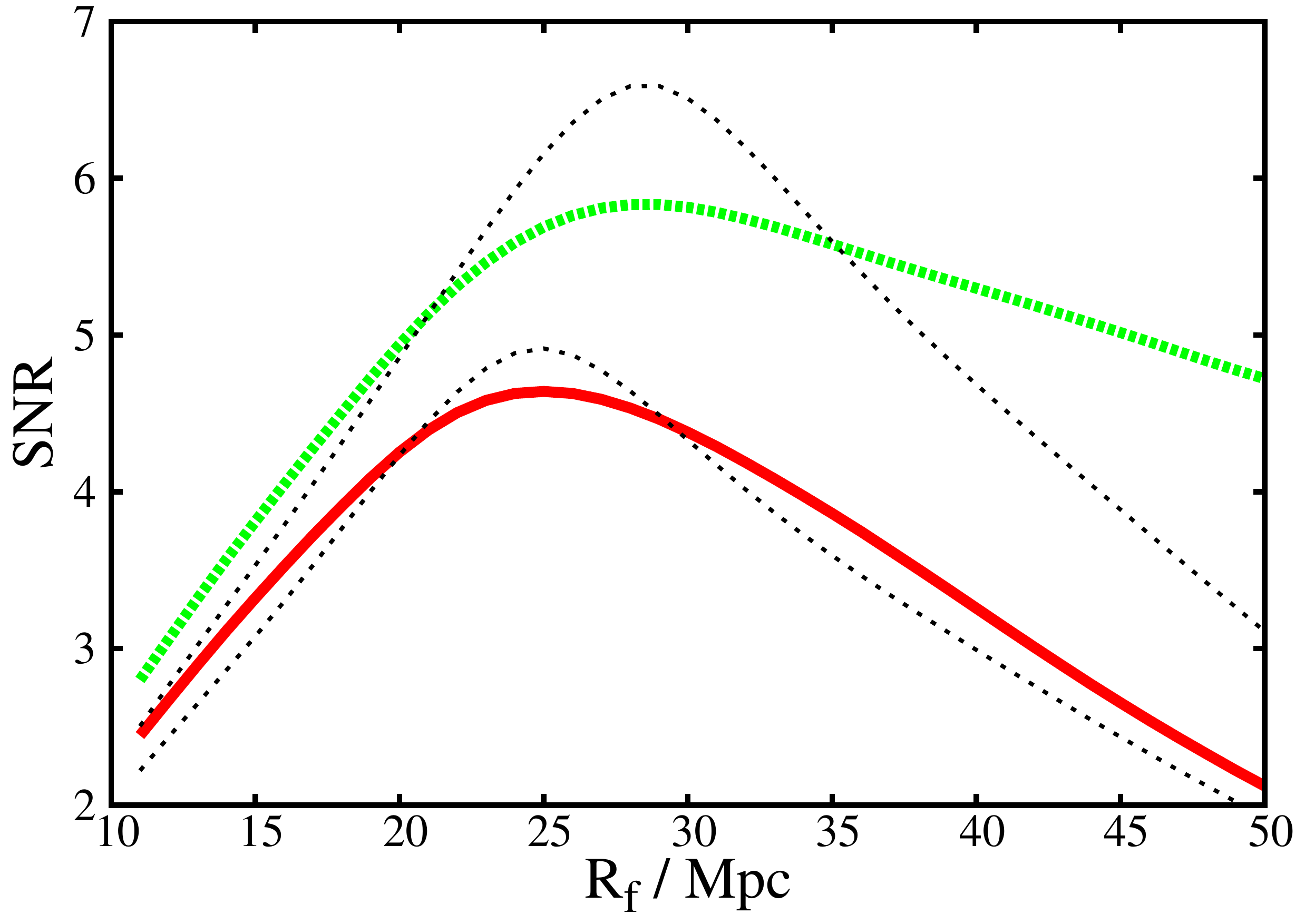}}
\caption{SNR=$\langle \hat{E} \rangle/\sqrt{\langle ( \Delta \hat{E})^2 \rangle_N}$ as a function of filter size $R_{\rm f}$ for $1200$ hours of LOFAR observations. Panel (a): the two curves (from top to bottom) represent the late simulations without the quasar ( see upper panels of Fig. \ref{fig:slices}b for the images) at two time steps corresponding to redshifts $7.57$ and $7.664$ respectively. Panel (b): same as panel (a) for the late quasar turn-on case with quasar (see lower panels of Fig. \ref{fig:slices} b for the images). Panel (c) shows the same for the large box simulation. The lower curve  shows for the quasar H II region (marked with `A'  in Fig. \ref{fig:dT_smooth}) surrounding the most massive DM halo. The upper curves shows the SNR for the adjacent H II region created by the stars only (marked with `B'  in Fig. \ref{fig:dT_smooth}). The two thin (black) lines are for spherical H II regions of radii $24.9 \, {\rm cMpc}$ (bottom) and $28 \, {\rm cMpc}$ (top) embedded in a uniform H I density at redshift 7.57. The lines are scaled to compare with the large box simulation results.}
\label{fig:snr-noqso1}
\end{figure*}

The filter $S_{\rm f} (\u , \nu)$  is defined as
\be
S_{\rm f} (\u , \nu)=S(\u , \nu)
\e
where $S_{\rm f}(\u , \nu)$ is the expected signal of the H II region of radius $R_{\rm f}$, at redshift $z_{\rm f}$ and spatial position $\theta_{\rm f}$.  According to the matched filter formalism, the SNR is maximized when  the filter parameters match exactly the H II region parameters i.e, $R_{\rm f}=R_{\rm b}$, $z_{\rm f}=z_{\rm b}$ and $\vec{\theta_{\rm f}}=\vec{\theta_{b}}$. Therefore, the peak in the SNR over $R_{\rm f}$ curve corresponds to H II region radius $R_{\rm b}$. We must note that the above filter is slightly different from the one we propose in  \citet{2007MNRAS.382..809D}. In  \citet{2007MNRAS.382..809D} (see Filter I), an extra term is introduced to subtract out the smooth foreground components. We drop this term here but re-include it later when we discuss the foreground subtraction in Sect. \ref{sec:foregrounds}. We also drop the overall factor $(\nu/\nu_{\rm c})^2$ from the filter. This term  accounts for the fact that $\rho_{ N}(\u,\nu)$ changes with frequency. As will be demonstrated in Sect. \ref{sec:foregrounds}, the residuals from the foreground subtraction are below the signal for LOFAR even if we drop this term from the filter.

The variance $\langle ( \Delta \hat{E})^2 \rangle$ of the estimator is a sum of three contributions
\be
\langle ( \Delta \hat{E})^2 \rangle=\langle ( \Delta \hat{E})^2 \rangle_{\rm HF}+\langle ( \Delta \hat{E})^2 \rangle_{\rm FG}+\langle ( \Delta \hat{E})^2 \rangle_{\rm N}, 
\e
where $\langle ( \Delta \hat{E})^2 \rangle_{\rm HF}$, $\langle ( \Delta \hat{E})^2 \rangle_{\rm FG}$ and $\langle ( \Delta \hat{E})^2 \rangle_{\rm N}$ are the contributions from H I fluctuations, foreground residual and the system noise, respectively. To calculate the variance from the H I fluctuations  $\langle ( \Delta \hat{E})^2 \rangle_{\rm HF}$ one would need to have many different realizations of the neutral density field surrounding always the same H II region. Simulating many EoR simulations of the kind we present here is computationally expensive and hence we do not discuss this here (see \citet{2008MNRAS.391.1900D} for approximate values of $\langle ( \Delta \hat{E})^2 \rangle_{\rm HF}$). Since the quasar H II region is much bigger than the average  size of surrounding H II regions, the contribution from $\langle ( \Delta \hat{E})^2 \rangle_{\rm HF}$) is expected to be much smaller than the signal $\langle\hat{E}\rangle$ \citep{2008MNRAS.391.1900D}.  The system noise variance $\langle ( \Delta \hat{E})^2 \rangle_{N}$ can be calculated analytically using the relation
\be
\langle ( \Delta \hat{E})^2 \rangle_{N}=\sigma^2 \int d^2 U \int d \nu \rho_N (\u, \nu)|S_{\rm f}(\u,\nu)|^2,
\label{eq:noise-rms}
\e
where $\sigma^2=\sqrt{\left [ \langle \hat{N^{\rm R}}^2 \rangle \right ]/ \Sigma_{a,b}}$. To test our noise simulations described in Sect. \ref{sec:noise_sim}, we calculate $\langle ( \Delta \hat{E})^2 \rangle_{\rm N}$ from 1000 independent noise realizations and  also using Eq. \ref{eq:noise-rms} for the same set of parameters and we find that they  agree with each other within $5 \%$. This test gives us the confidence that our noise simulations are correct. We use the simulated noise to further predict the constraints LOFAR should be able to  put on different H II region sizes. We will discuss this in Sect. \ref{sec:noise}. 

\section{Results of the Matched Filter Technique}
\label{sec:results2}
Here we discuss the detectability of the quasar H II regions in our simulations for LOFAR. Next we discuss if LOFAR can identify quasar H II regions blindly. We discuss the effect of system noise on the size determination of an H II region and the accuracy in size determination we can achieve for LOFAR.  

\begin{figure*}
\labellist
\small\hair 2pt
\pinlabel $\rotatebox{90}{\textbf{(a)}}$ at 31 647 
\endlabellist
\rotatebox{270}{\includegraphics[clip,height=.95\columnwidth]{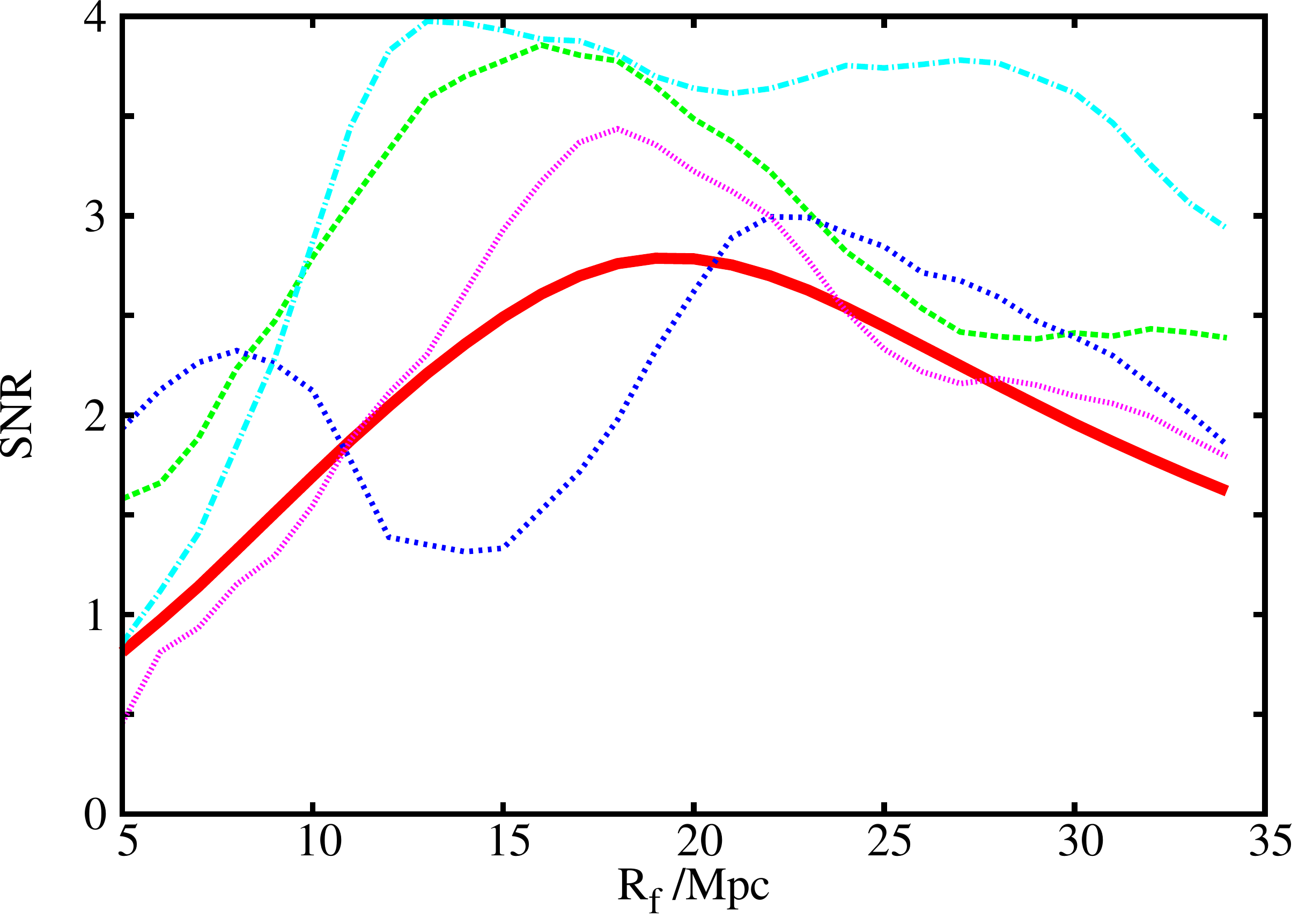}}
\labellist
\small\hair 2pt
\pinlabel $\rotatebox{90}{\textbf{(b)}}$ at 31 647 
\endlabellist
\rotatebox{270}{\includegraphics[clip,height=.95\columnwidth]{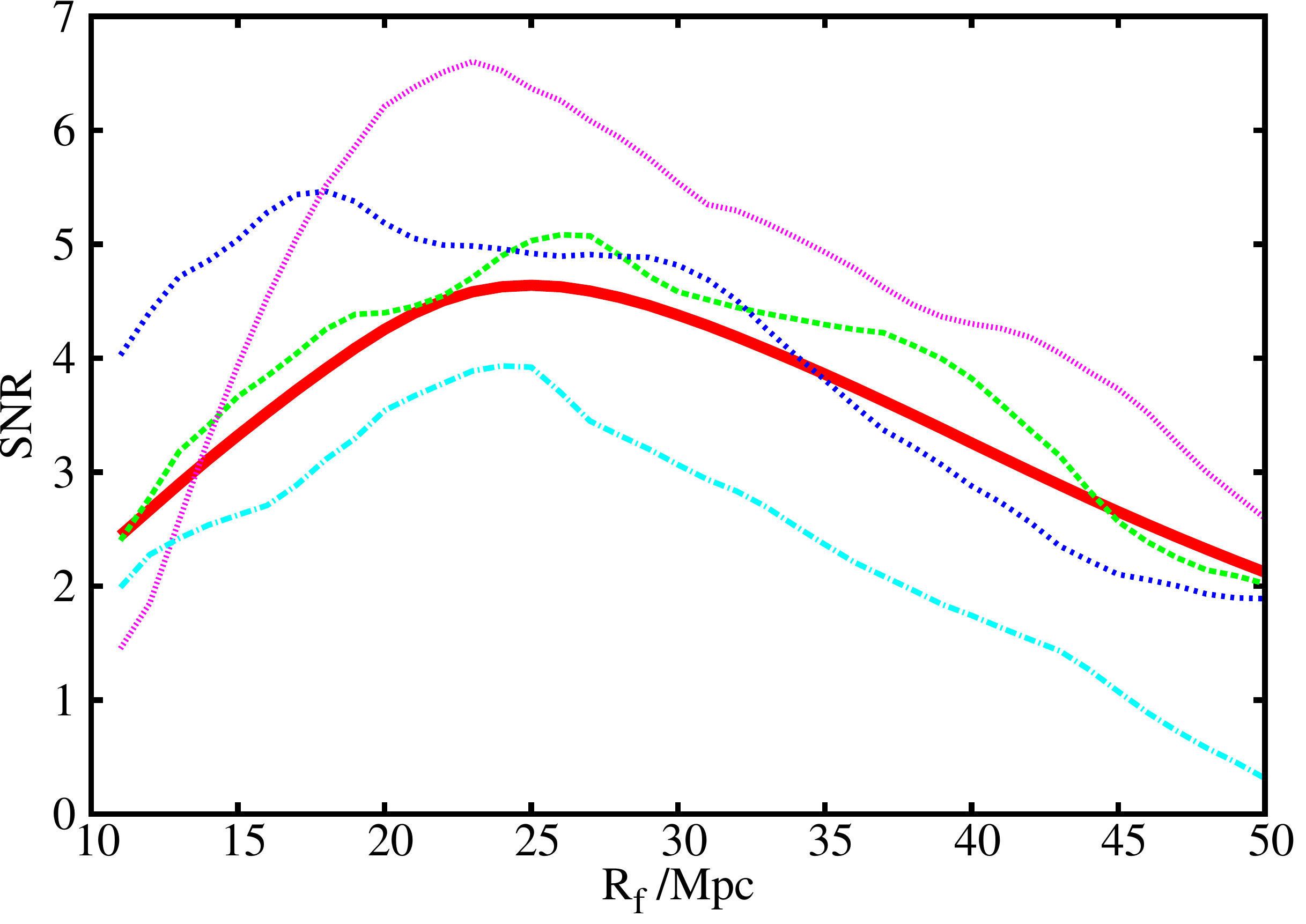}}
\caption{The effect of adding the system noise to the signal. The left panel (a) shows results for the late simulation with quasar for $1200$ hours of LOFAR observations. Thick solid line shows SNR for different filter sizes $R_{\rm f}$ for the signal only. Other lines represent randomly selected cases when we add different noise realizations to the signal.  Right panel (b) shows the same for the large box simulation with quasar.}
\label{fig:noise+sig}
\end{figure*}
\subsection{Detectability}

Fig. \ref{fig:snr-noqso1} shows SNR$=\langle \hat{E} \rangle /\sqrt{\langle ( \Delta \hat{E})^2 \rangle_{\rm N}}$ as a function of filter size $R_{\rm f}$ for our different simulations for $1200$ hours of LOFAR observations. Note that here we do not add system noise to the signal but calculate the expected noise rms $\sqrt{\langle ( \Delta \hat{E})^2 \rangle_{\rm N}}$ using Eq. \ref{eq:noise-rms}. In Sect. \ref{sec:noise} we discuss in detail the effect of noise when we add noise to the signal. Here we take the filter position $(\th_{\rm f}, \nu_{\rm f})$ to be the same as the H II region position $(\th_{\rm b}, \nu_{\rm b})$. This corresponds to a targeted observation i.e, the central quasar's position and redshift are known from an  optical/near infra-red survey and we point the radio telescope to it (see \cite{2008MNRAS.391.1900D} for blind search). The bottom line (red) in the right panel (c) shows the result for the quasar H II region in the large box simulation. We see that the SNR peaks at filter size $R_{\rm f}=24.6 \, \rm{cMpc}$ and the peak SNR is $4.6$.  According to the matched filter principle the SNR peaks when the filter matches best with the signal. Therefore  the above $R_{\rm f}$ value is interpreted as the H II region size found by our method. We also calculate the expected H II region size in the simulation based on the total number of ionizing photons (see Sect.~\ref{sec:results1} ) which gives $25.1 \, \rm{cMpc}$. We see that both agree well with each other. The  middle panel of Fig.~\ref{fig:snr-noqso1} (b) shows results for the late quasar turn-on case. The two different lines correspond to two different times after quasar turn on. The left panel (a) is the same as the middle but without the quasar. The peak SNR, H II region radii obtained both from matched filter method and from total photon counts for different simulations are summarized in Table \ref{table:results-matched}. We see that LOFAR has good prospects to detect our quasar H II regions at redshift $7.57$ when the global ionization fraction is $\sim 50 \%$. As the LOFAR FoV is bigger than our simulation box size, the chance of having at least one such detectable quasar H II region is very good. We also see that the H II region sizes obtained from the matched filter method are in good agreement with the expected H II region sizes from total ionizing photon counts. Therefore we conclude that the H II region size obtained from the matched filter method can be used to calculate the total number of ionizing photons within a spherical region of size equal to the matched H II region size. This can be further used to constrain the quasar ionizing photon rate and age. We discuss this in Sect. \ref{sec:summary}

\begin{table}
\caption{Summary of results obtained from the matched filter method.}

\resizebox{\columnwidth}{!}{
\begin{tabular}{l  c c c c c }

\hline
 & quasar age	& z &Peak SNR  &  H II region size &  H II region size  \\
 &             &    &        &  (from filter)&   (from total photon) \\
                  \hline \\[-6pt]
early quasar & 23.0 Myr & 8.397 &1.2    & 11.6 cMpc   &  12.0 cMpc \\
late quasar  & 11.5 Myr & 7.664 &2.2 & 16.0 cMpc &  16.4 cMpc  \\
          & 23.0 Myr & 7.570 &2.8 & 19.4 cMpc &  19.7  cMpc\\
large box & 23.0 Myr & 7.570 &4.6 & 24.9 cMpc & 25.1 cMpc \\[1pt]
\hline

\end{tabular} }
\label{table:results-matched}
\end{table}

The upper (green) thick curve in the panel (c) in Fig. \ref{fig:snr-noqso1} represents the large H II region very close to the quasar H II region in our large box simulation (marked with `B' in Fig. \ref{fig:dT_smooth}, see also Fig. \ref{fig:dT_3D}, panel b). We discussed in Sect. \ref{sec:results1} that the H II region is bigger than the one around the luminous quasar. We find the H II region size to be $\sim 28.7 \, {\rm cMpc}$ and the peak SNR achievable by LOFAR for $1200$ hours of observations is $5.8$. Therefore, we argue that LOFAR cannot alone distinguish between H~II regions made by clustered galaxy sources and H~II regions made by a luminous quasar by measuring the size of the H II region. Another possible way could be to use the shape of the SNR vs $R_ {\rm f}$ curve to extract the information about the sphericity of an H~II region. The SNR vs $R_ {\rm f}$ curve will be more peaked for a spherical H II region than an a-spherical one. The two thin (black) lines in the panel (c) of Fig. \ref{fig:snr-noqso1} show the results for spherical H II regions of radii $24.9 \, {\rm cMpc}$ (bottom thin line) and $28 \, {\rm cMpc}$ (top thin line) embedded in a uniform H I density at redshift 7.57. The lines are
scaled to compare with the large box simulation H II regions around quasar and stellar sources (marked with A and B in Fig. \ref{fig:dT_smooth}, respectively). As expected the SNR vs $R_ {\rm f}$ curves for spherical H II regions are more peaked than the simulation. But  we note that the quasar H~II region is more spherical than the stellar H II region. So in principle the shape of the SNR vs $R_ {\rm f}$ carries information about the shape of the H II regions. However, as described below in Sect. \ref{sec:noise}, the system noise will make it impossible to use the shape of the SNR-$R_{\rm f}$ curve to make the statement about the sphericity of the regions found. 
 
Next we discuss the effect of the size of the smaller box that we use here. The size of the early and late simulation box is $\sim 1^{\circ}$ at the redshifts of our interest. This is much smaller than the LOFAR FoV ($\sim 5^{\circ}$). The minimum baseline we sample for the early and late simulations  is $U_{\rm min}\sim 56$ for this smaller box. A substantial amount of signal comes from the large scales for the H II regions we are interested in. In addition, LOFAR has many small baselines (see right panel of Fig. \ref{fig:coordinate}), so we lose a considerable amount of the signal in the estimator. The amount of loss depends on the H II region size: for larger H II regions the loss is higher than for smaller. We use our analytical equations to calculate the loss in the peak SNR for different H II region sizes. For H II regions of radius $20,\, 30$ and $40\, {\rm cMpc}$ we lose $\sim 3 \%$, $\sim 4 \%$ and $\sim 7 \%$ in the SNR, respectively. For the large box this will be even smaller and hence we do not discuss it here.  

\begin{figure}
\includegraphics[width=.52\textwidth, angle=0]{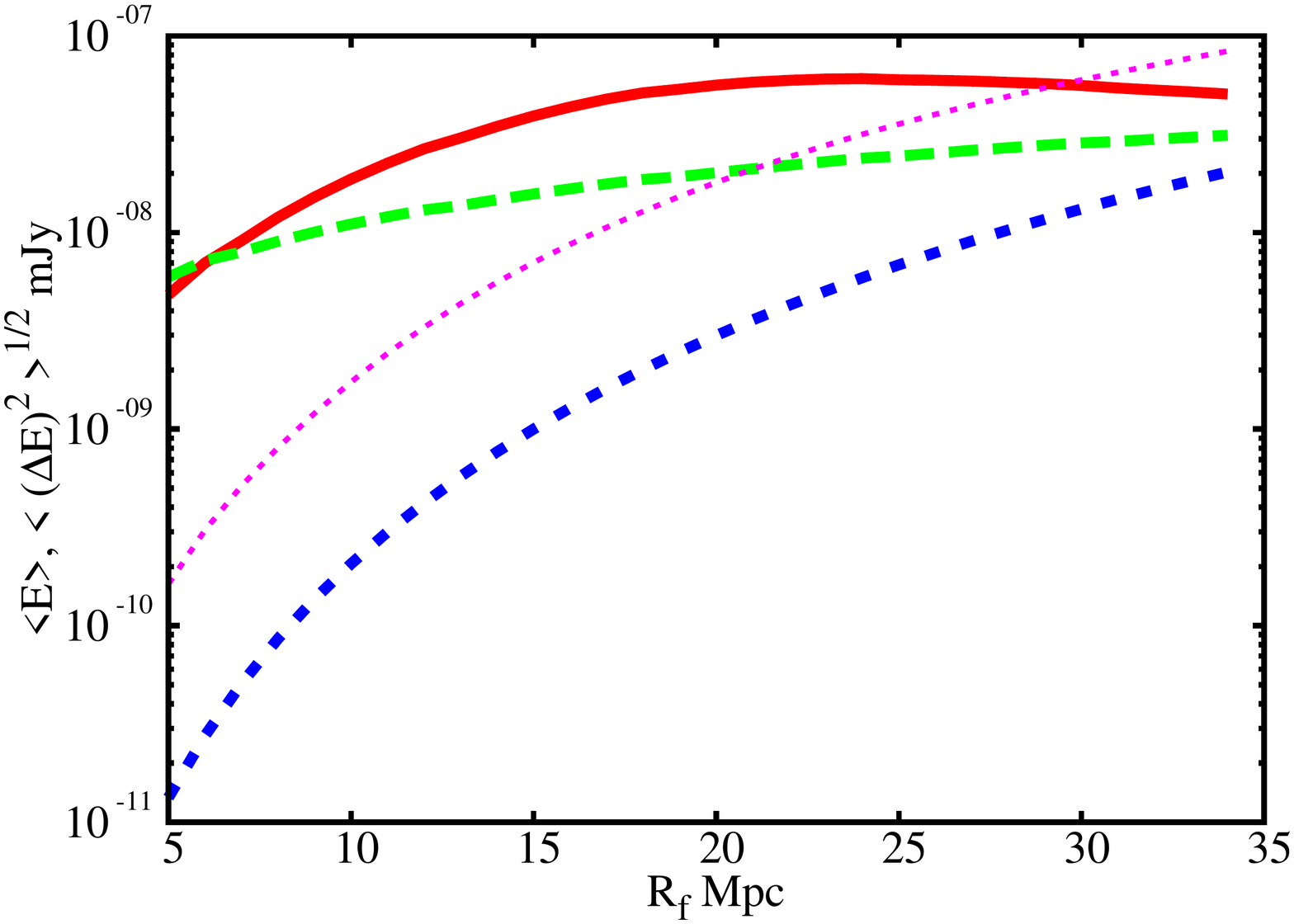}
\caption{The estimator $\langle E\rangle$ (solid thick red line) for the quasar late simulation is compared to the system noise rms $\sqrt{\langle(\Delta E)^2 \rangle_{\rm N}}$ (green dashed line) for $1200$ hours LOFAR observations and the foreground residual $\sqrt{\langle(\Delta E)^2 \rangle_{\rm FG}}$ (magenta and blue dashed lines) for the bandwidth $6$ times the H II region size in frequency.  The magenta and blue dashed lines show for the cases when $\rho(U,\nu)$ is frequency dependent and independent respectively (see text for details). }
\label{fig:fg-resi}
\end{figure}

\subsection{Effect of System noise on the size determination}
\label{sec:noise}
In the previous section we discussed results without explicitly adding system noise to the simulated signal, but  only calculated the expected noise rms in the estimator. Here, we add the simulated noise realizations to the signal and then perform the matched filtering. The left panel (a) of Fig. \ref{fig:noise+sig} shows the SNR$=\langle \hat{E} +\hat {\Delta E}_{\rm N} \rangle /\sqrt{\langle ( \Delta \hat{E})^2 \rangle_{\rm N}}$ vs $R_{\rm f}$ for four randomly selected noise realizations for $1200$ hours of LOFAR observations for the late quasar simulation. The solid thick (red) line shows the result without the noise. We see that in some cases there are clear peaks but shifted to either side of the original peak. We also see various features arising because of the noise such as a nearly flat SNR curve (no obvious peak), double peaked, suppressed and enhanced SNR.  In real observations one of these SNR curves will be observed. Now we only consider the realizations where the peak SNR is higher than $3$ (because a SNR less than 3 is not considered a detection at all), and then find the best matched filter size $R_{\rm f}$ from the peak SNR. We get an average H II region size $\langle R_{\rm f} \rangle=19 \,{\rm cMpc}$ and $\sigma_{R_{\rm f}}=4.06 \,{\rm cMpc}$.  The right panel shows the results for the large box simulation. In the simulation the H II region is detectable with SNR=$4.6$. As expected, the peaks are more prominent in almost all cases and therefore the H II region size can be detected with higher accuracy with $\langle R_ {\rm f} \rangle=25.3 \, {\rm cMpc}$ and $\sigma_{R_ {\rm f}}=4.0 \, {\rm cMpc}$. H II region size can be measured more accurately for higher SNR measurements. We also find that $\sigma_{R_{\rm f}}=2.5$, $1.75$, $1.2 \,{\rm cMpc}$ for our late quasar simulation where the H II region radius is $19 \, {\rm cMpc}$ if we have  $5\sigma$, $7\sigma$ and $10 \sigma$ detection respectively.
\\

\section{Foreground subtraction and its effect on the detection of a H II region}
\label{sec:foregrounds}

Astrophysical foregrounds are expected to be several orders of magnitude stronger than the H I signal \citep{2008MNRAS.385.2166A, 2009A&A...500..965B,  2011MNRAS.413.1174P, 2005ApJ...625..575S, 2008MNRAS.389.1319J} at frequencies relevant for EoR studies. Major components are the diffuse synchrotron and free free emission from our own Galaxy, extra galactic point sources etc. \citep[see][ for more detail]{2008MNRAS.389.1319J}. Various methods have been developed to subtract the foregrounds \citep{2006ApJ...650..529W, 2008MNRAS.389.1319J,  2009MNRAS.397.1138H,   2009ApJ...695..183B, 2009MNRAS.398..401L,  2011MNRAS.411.2426G, 2011MNRAS.418.2584G, 2012ApJ...744...29M, 2012arXiv1201.2190C}). All the methods rely on the fact that the foreground components are smooth in frequency. As we first proposed in  \citet{2007MNRAS.382..809D}, we modify the matched filter to subtract the foreground below the signal. The modified filter is,

\be
S^{\rm M}_{\rm f} (\u , \nu)=S(\u , \nu)-\frac{1}{\rm BW} \int^{\nu_{\rm f}+ {\rm BW}/2}_{\nu_{\rm f}-{\rm BW}/2}S(\u,\nu) d\nu
\e
where BW is the bandwidth over which the foreground subtraction is done. $\nu_{\rm f}$ is the frequency centre of the filter. Note that $\int^{\nu_{\rm f}+{\rm BW}/2}_{\nu_{\rm f}-{\rm BW}/2}S^{\rm M}_{\rm f}(\u,\nu) d\nu=0$ and $S^{\rm M}_{\rm f}(\u,\nu)$ is also symmetric around the frequency centre $\nu_{\rm f}$ i.e, $S^{\rm M}_{\rm f}(\u,\nu-\nu_{\rm f})=S^{\rm M}_{\rm f}(\u,-(\nu-\nu_{\rm f}))$. This means, the filter will subtract any frequency independent component in the frequency range $\nu_{\rm f}-{\rm BW}/2$ to $\nu_{\rm f}+{\rm BW}/2$. Because of the symmetry, the filter will also subtract out any component which is  an odd function of  $(\nu-\nu_{\rm f})$ i.e,   components proportional to $(\nu-\nu_{\rm f})$ (linear),  $(\nu-\nu_{\rm f})^3$ and so on. However the quadratic terms will remain as the residuals.  

We next calculate the contribution of  foreground residuals to the variance of the estimator using the the analytical formula (see \cite{2007MNRAS.382..809D} for more details)
\begin{eqnarray}
\langle ( \Delta \hat{E})^2 \rangle_{FG} &=& \int d^2 U \int d\nu_1 \int d\nu_2  \left ( \frac{dB_{\nu1}}{dT}\right) \left ( \frac{dB_{\nu2}}{dT}\right)  \nonumber  \\
&  & \times \rho(\u,\nu_1) \rho(\u,\nu_2) {S^{\rm M}}^{*}_{\rm f}(\u,\nu_1)S^{\rm M}_{\rm f}(\u,\nu_2) \nonumber \\
&  & \times C_{2 \pi U} (\nu_1, \nu_2)
\label{eq:fg-residual}
\end{eqnarray}
where $\left ( \frac{dB_{\nu}}{dT} \right )$ converts the temperature to the specific intensity and $C_{2 \pi U} (\nu_1, \nu_2)$ is the multi-frequency angular power spectrum of all foreground components. 

For each component of the foreground the multi-frequency angular power spectrum is modelled as
\be
C_{2 \pi U}(\nu_1, \nu_2)=A \left ( \frac{\nu_{\rm f}}{\nu_1} \right )^{\bar{\alpha}} \left ( \frac{\nu_{\rm  f}}{\nu_2} \right )^{\bar{\alpha}} \left ( \frac{1000}{2 \pi U} \right )^{\bar{\beta}}. 
\label{eq:fg-model}
\e
We assume the foreground components to be correlated at two different frequencies $\nu_1$ and $\nu_2$ and ignore possible small decorrelations. We consider two major foreground components: diffuse galactic synchrotron emission from our own galaxy and point source contribution (clustered and Poisson part). We have adopted the parameter values for $A$, $\bar{\alpha}$ and $\bar{\beta}$ from \citet{2005ApJ...625..575S}. Under the assumption of the above foreground modelling  Eq. \ref{eq:fg-residual} can be written as 
\bear
\langle ( \Delta \hat{E})^2 \rangle_{\rm FG} &=& 2 \pi \sum_i A_i \int_{U_{\rm min}}^{U_{\rm max}} dU \, U \left ( \frac{1000}{2 \pi U} \right)^{\beta_i} \nonumber \\
&&\left [ \int_{\nu_{\rm f}-{\rm BW}/2}^{\nu_{\rm f}+{\rm BW}/2} d \nu  f_i(U, \nu) |S^{\rm M}_{\rm f}(U, \nu)| \right ]^2 
\label{eq:fg-residual1}
\ear
where the index `i' indicates different foreground components and  $f_i(U, \nu)=\left ( \frac{dB_{\nu}}{dT} \right )  \rho(U, \nu) \left ( \frac{\nu_{\rm f}}{\nu} \right )^{\bar{\alpha_i}}$. Note that here we assumed the baseline distribution $\rho(\vec{U},\nu)$ to be circularly symmetric. This assumption together with the foreground model we assumed in Eq. \ref{eq:fg-model} reduces  Eq. \ref{eq:fg-residual} from four integrals to two and hence the evaluation of Eq. \ref{eq:fg-residual1} is very fast with very good accuracy. The terms $\left ( \frac{dB_{\nu}}{dT} \right )$, $\rho(U, \nu)$, $\left ( \frac{\nu_{\rm f}}{\nu} \right )^{\bar{\alpha_i}}$ are expected to be smooth in frequency. Therefore they can be written as (in the frequency range $\nu-{\rm BW}/2<\nu<\nu+{\rm BW}/2$)
\bear 
f(U, \nu)&=&f(U,\nu_{\rm f})+(\nu-\nu_{\rm f})f'(U,\nu_{\rm f}) \nonumber \\
&+&(\nu-\nu_{\rm f})^2\frac{f''(U,\nu_{\rm f})}{2!} \nonumber \\
&+&(\nu-\nu_{\rm f})^3\frac{f'''(U,\nu_{\rm f})}{3!}+...  
\ear
where the prime denotes derivatives with respect to $\nu$. As we have discussed before in this section, the modified matched filter $S^{\rm M}_{\rm f}(U, \nu)$ removes the constant part $f(U,\nu_{\rm f})$, the linear term $(\nu-\nu_{\rm f})f'$ and all terms with odd powers. Only the quadratic term $(\nu-\nu_{\rm f})^2\frac{f''(U,\nu_{\rm f})}{2!}$ and other terms with even powers remain as a residual. This is the reason why the filter is so successful in removing the foreground below the signal.

In Fig. \ref{fig:fg-resi} we plot $\sqrt{ \langle ( \Delta \hat{E})^2 \rangle_{\rm FG}}$ for different filter sizes $R_{\rm f}$ (magenta dashed line) for the bandwidth $6$ times the filter size $R_{\rm f}$ in frequency. For comparison we also plot the estimator $\langle \hat{E} \rangle$ for our smaller box simulation with quasar at $z=7.57$ and the noise rms $\sqrt{ \langle ( \Delta \hat{E})^2 \rangle_{N}}$. We see that $\sqrt{ \langle ( \Delta \hat{E})^2 \rangle_{\rm FG}}$ (magenta dashed line) is smaller than the estimator for the filter size $R_{\rm f}< 30 \, {\rm cMpc}$.  We further find that a significant amount of the residual foreground comes from the fact that the baseline distribution $\rho(U,\nu)$ is frequency dependent i.e, the number of baselines in a uv cell changes with frequency. Therefore the effective foreground which basically depends on the product $\rho(U,\nu)\nu^{-\bar \alpha}$ (see Eq. \ref{eq:fg-residual1}) becomes less smooth than the original foreground. Since the baseline distribution at different frequencies is completely known, the change in the number of baselines with frequency for a uv cell can be compensated by introducing this in the filter. This way  the effect of $\rho(U,\nu)$ changing with frequency can be avoided. The blue dashed line in Fig.  \ref{fig:fg-resi}  shows $\sqrt{ \langle ( \Delta \hat{E})^2 \rangle_{\rm FG}}$ for the baseline distribution independent of frequency. We see that the residual is much below the estimator for the case where $\rho(U,\nu)$ is frequency independent. 

We note  that the modified filter subtracts out some H I signal as well. The amount of H I signal it subtracts depends on the size of the bandwidth BW over which the foreground subtraction is done. The larger the bandwidth, the smaller is the loss of signal. For H~II regions of radii between $20 \, {\rm cMpc}$ to $25\, {\rm cMpc}$ at redshift $7.57$ the loss in  SNR is around ~$24\%$, $15\%$ and $10\%$ for the BW $4$, $6$ or $8$ times the filter size $R_{\rm f}$ in frequency. On the other hand, the level of foreground residuals increases for larger bandwidths. The choice of optimal bandwidth depends on the size of H II region and the foreground model. In our case  we find that a bandwidth $6$ times the radius of H II region  in frequency is reasonable. We notice that the modified matched filter does not peak at the H II region radius  but at a slightly large size. We find that the SNR peaks at filter radius $R_{\rm f}=25.4 \, {\rm Mpc}$ for a spherical H II of radius $R_b=25 \, {\rm Mpc}$ at redshift $7.57$ for a bandwidth $6$ times the H II region radius. This is a small effect compared to the uncertainty from  the system noise. 

\section{Discussion and Conclusions}
\label{sec:summary}

We investigated three cases of a luminous quasar shining during the epoch of reionization in the most massive halos of our simulation volumes. Two cases used a smaller simulation volume with a side length of 163 comoving Mpc and one used a substantially bigger volume with a side length of 607 comoving Mpc. One of the cases in the smaller volume considered a quasar turn on at an early stage of reionization, when the global ionization fraction was about 25\%. The other two cases considered a quasar turn on at a later stage of reionization, when the global ionization fraction was about 45\%. We found that the 163 Mpc simulation volume was essentially too small to be able to host many large scale H~II regions at the same time and therefore, we concentrated on the larger simulation volume to investigate if the quasar changes the size and the morphology of the H~II region in a way which makes the quasar H~II region distinctly different from H~II regions solely formed by clustered galaxy sources. We found that although the quasar dominates the ionizing photon output in, and the growth of the H~II region hosting the quasar, other highly overdense regions with many clustered massive halos can produce H~II regions similar or even larger in size than the H~II region hosting the quasar. Although detailed investigation of the H~II regions shows that the extend in all 3 spatial dimensions of the quasar H~II region is more equal (i.e. the quasar H~II region is more spherical) than large H~II regions by galaxies alone, given observational constrains, it might not be possible to tell the difference using 21cm observations. Furthermore, we found that using a quasar spectrum $L\propto \nu^{-1.5}$, the sharpness of quasar H~II regions will not be measurably different from galaxy H~II regions.

Next we use the matched filter technique, which combines the signal optimally, to investigate the possibility of detecting a single QSO HII region in redshifted 21 cm observations using LOFAR. We find that with 1200 hrs of observations, LOFAR should be able to detect a single QSO HII region of comoving radius 20 Mpc and 25 Mpc or bigger with at least 3 and 5 $\sigma$ significance, respectively at redshift 7.57 when the universe is 50\% ionized. We also find that the H~II region formed by a quasar at earlier stages of reionization, in our early quasar turn on case, is not large enough to be detectable. 
We calculate the accuracy at which the bubble size can be measured in case of a significant detection. For our best case, the quasar in the 607 Mpc simulation, we find for 100 realisations of including system noise $R_{\rm b}= (25.3 \pm 4.0) \, \rm{Mpc}$ assuming 1200 hours of LOFAR integration time.

We also considered the effect of foregrounds. For this,
we modified the matched filter to subtract the foreground components. For a reasonable foreground model the modified matched filter is able to subtract the foregrounds such that the residuals are far below the signal level.

Since we found that clustered galaxy sources in highly overdense regions are able to form H~II regions of comparable size to the quasar H~II region, we also used the matched filter technique on a non-quasar H~II region to investigate if the form of the SNR-curve can be used to tell quasar H~II regions apart from H~II regions solely formed by clustered galaxy sources. Although the resulting SNR-curves are substantially less peaked, the system noise hides this information. We therefore conclude that a blind search for quasars using redshifted 21cm maps is not likely to be successful. 

The radii $R_{\rm b}$ found by the matched filter method for the three cases after 23.0 Myr are 
11.6 cMpc, 19.4 cMpc and 24.9 cMpc. This is in rough agreement with estimates for the expected bubble size on the basis of photon counts, $R=$ 12.0 cMpc, 19.7 cMpc and 25.1 cMpc taking into account both the stellar photons emitted in the quasar region and the quasar photons but neglecting recombinations. We can now use the radii found by the matched filter method and Eq.\ref{eq:eq1} to calculate backwards to estimate the rate of ionizing photons emitted by the quasar.
Since the best prospects of detecting an H~II region by the matched filter method are for the quasar in the large simulation volume, we restrict the following analysis to this case.

Without other information on the quasar, we can assume that the minimum time $t_{\rm on}$, the quasar has been shining at the time of observation corresponds to the light-travel time from the centre to the edge of the H~II region, so $t_{\rm on}\gtrsim t_{\rm on}^{\rm min} = R_{\rm b}/(1+z)/c$, where $c$ is the speed of light. 

Using this lower limit results in an upper limit for the ionizing photon rate from the quasar. Rewriting Eq.\ref{eq:eq1} to include an unknown contribution of stellar photons $N_{\gamma}^*$ (in terms of total emitted photons contributing to the H~II region), the ionizing photon rate from the quasar can be estimated to
\bq
\dot{N}_{\gamma}^{\rm qso}\lesssim \left(R_{\rm b}^2\frac{4}{3}\pi n_c-\frac{N_{\gamma}^*}{R_{\rm b}}\right)c \, (1+z)
\eq
Here, $n_{\rm c}$ is the comoving average number density of atoms in the universe. Assuming  $n_{\rm c}$ as average number density in the H~II region is a good estimate for radii larger than about 20 cMpc. 

If we now concentrate on our best case, the 607 cMpc simulation, we get the following result:
\bq
\dot{N}_{\gamma}^{\rm qso} \lesssim (1.26 \times 10^{57}- 3.34 \times 10^{-15} \, N_{\gamma}^*) \, \rm{s}^{-1} \, .
\eq
If we neglect the contribution from the clustered galaxies completely, this gives $\dot{N}_{\gamma}^{\rm qso} \leq 1.26 \times 10^{57}$ s$^{-1}$. 
This is roughly a factor 5 more than the ionizing photon rate that was used in the simulation. Roughly, a factor 2 comes from neglecting the contribution from the clustered galaxies, a factor 2.4 comes from the time estimate. The 1$\sigma$ uncertainty from the size estimate translates into $\sigma(\dot{N}_{\gamma}^{\rm qso})=4.0 \times 10^{56}$~s$^{-1}$. 

In case the rate of ionizing photons is known by estimates of the total luminosity and assumptions about the unabsorbed continuum spectrum \citep[such an estimate was made for the $z=7.085$ quasar by][ and resulted for that case in an ionizing photon rate of $\Gamma_{\rm ion} = 1.3 \times 10^{57}$ s$^{-1}$]{2011Natur.474..616M}, the radii found by the matched filter method can be used to put 
lower limits on the quasar lifetime. Assuming we would be able to find the exact values for the photon rates, see Table \ref{table:summary2}, we would estimate the quasar on time to be: 
\bq
t_{\rm on}=\frac{R_{\rm b}^3 \frac{4}{3}\pi n_c-N_{\gamma}^*}{\dot{N}_{\gamma}^{\rm qso}}
\eq
Ignoring the contribution from the galaxy clusters $N_{\gamma}^*$ results in 49 Myr. This is roughly a factor of two longer than the actual quasar on time, as would have been expected given the total photon numbers from the quasar and the clustered galaxy sources, as presented in Table \ref{table:summary2}. The 1$\sigma$ uncertainty from the size estimate translates into $\sigma(t_{\rm on})=23$~Myr.
However, given that current uncertainties of quasar ages are higher than this\footnote{\citet{1998MNRAS.300..817H} find quasar lifetimes to be $10^6 \-- 10^8$ yr from matching stellar and quasar luminosity functions at $z\sim 3$.
\citet{2001ApJ...547...27H} also find quasar lifetimes  around $10^6 \-- 10^8$ yr from clustering analysis of dark matter halos, modelling luminosity function and matching with clustering of quasars. 
\citet{2011ApJ...736...49L} discuss that recent studies of the proximity effect using Ly$\alpha$ forest spectra seem to be contradictory since some imply lifetimes $\leq 10^6$ yr \citep[e.g.][]{2008MNRAS.391.1457K} while other imply lifetimes $\geq$ a few $10^7$ yr \citep[e.g.][]{ 2006A&A...450..495W}.}, it would still be an improvement for limiting the quasar lifetime. As \citet{2011arXiv1111.6354M} already point out in their conclusions, the main advantage of using the matched filter method is, that one does not need to rely on a few line-of-sight measurements but instead, takes the full 3D information into account.

Obviously our study is limited to the cases we considered. We ignore any relativistic effects here, both in simulations and in the estimates. The relativistic effects result in smaller H II regions than the case where they are not included. It can be quite significant for the initial phases of the expansion of an H II region around a luminous QSO, but becomes less important as time progresses \citep{2006ApJ...648..922S}. However, we expect the effect to be small for our cases for two reasons: First,  the ionizing photon luminosity ${\dot N}^{\rm qso}_{\gamma}$ even for the brightest QSO in the large box simulation is a factor $\sim 4$ lower than considered ($10^{57}/s$) in \citet{2006ApJ...648..922S} and secondly, the QSO turns on in a big H II region already created by stellar sources. Our study is limited to the quasar luminosities we assumed.  If the quasar luminosity would be substantially higher, the H~II region it is
located in would be substantially larger. This
would increase their detectability with the matched filter technique
and would increase the relative accuracy of the estimate of their
total ionizing photon production and life times. More luminous quasars
than we considered could occur for two main reasons, namely sample
variance and a higher conversion factor between halo mass and quasar
luminosity.

Even though the size of our largest simulation volume approaches that
of a single LOFAR field, the entire sky presents a volume $\sim 1000$
larger and more massive halos
may exist. However, such massive halos will live in extremely biased
regions, corresponding to high $\sigma$ peaks in the density, with
associated high levels of ionized flux from galaxies. Our results
suggest that, for our assumed efficiency parameters for both galaxies
and quasars, the quasar contribution to the total number of photons
ionizing that region will be around 50\%, with a {\it decreasing}
trend for higher $\sigma$-peaks. Consequently, although the HII region
around a more massive halo should be larger and therefore easier to
detect in the 21cm observations, it will be considerably affected by
stellar photons, which will have to be taken into account when
analysing its size. In addition, if the trend we see is correct, in a
relative sense the quasar may contribute less to such a region, and
therefore have a lesser impact on it.

A higher luminosity efficiency for our halos would be possible if
black hole growth is substantially more efficient than we assumed.
However, for a reasonable Eddington efficiency $\lambda$, we are
already close to the upper limit of the halo mass -- black hole mass
relation which assumes all baryons in the galaxy to be located in the
bulge. Also, the most massive black hole found at $z=7.5$ in the
simulations of \citep{2012ApJ...745L..29D}, who followed the black
hole growth in detail in a volume comparable to our large box, is not
that much more massive than what we assume. So, it would seem to
require a strong discrepancy between the joint growth of the halo and
its black hole in order to obtain quasars that are an order of
magnitude more luminous than we have assumed.

The quasar could also have a larger impact if the ionizing photon
output from galaxies would be lower than we have assumed. However,
current constraints on reionization already limit the efficiency to
factors of a few around the values assumed here. Much lower
efficiencies would not only take us further away from the electron
scattering optical depth as measured by WMAP but also result in a
completion of reionization after $z\sim 6$. In fact, since we did not
take into account gas clumping below the resolution of our radiative
transfer grid nor (at least in the smaller of the two volumes
concerned) the absorbing effect of Lyman Limit Systems, slightly
higher ionizing efficiencies for the galaxies could actually be needed
for reionization to complete before $z\sim 6$. In any case, one may
hope that a detection of the redshifted 21cm signal over a substantial
redshift range will give us an approximate reionization history which
could then be used to constrain the ionizing photon output from
galaxies.

As we have shown, the largest HII regions at a given epoch do not
necessarily host a quasar and so the detection of such a region in
21cm observations cannot be used as a secure method for finding
quasars. However, since such a region does point to an overdense part
of the Universe and they are relatively rare, it would still make
sense to use such them for follow up deep optical/near-infrared
searches for quasars since they may indeed contain one. Although an
exploration of the efficiency of processing redshifted 21cm data using
a full four-dimensional (position and size) matched filter search is
beyond the scope of this paper, our results do seem to suggest this to
be a feasible approach.

\section*{Acknowledgments} 
This study was supported in part by the Swedish Research Council grant 2009-4088. 
The authors acknowledge the Swedish National Infrastructure for Computing (SNIC) resources at HPC2N (Ume\aa, Sweden) and PDC (Stockholm, Sweden) and TeraGrid and the Texas Advanced Computing Center (TACC) at The University of Texas at
Austin ( URL: http://www.tacc.utexas.edu) for providing HPC resources. The authors also acknowledge the Royal Society International Join Project grant. The authors would like to thank Michiel Brentjens for providing us with the LOFAR station coordinates and Kyungjin Ahn for providing the recipe for including sub-resolution sources in the 504 Mpc simulation volume.
KKD is grateful for financial support from Swedish Research
Council (VR) through the Oscar Klein Centre (grant 2007-8709).
ITI was supported by The Southeast Physics Network (SEPNet) and the Science and Technology Facilities Council grants ST/F002858/1 and ST/I000976/1.
PRS was supported by  NSF grants AST-0708176 and AST-1009799, NASA grants NNX07AH09G, NNG04G177G and NNX11AE09G, and Chandra grant SAO TM8-9009X.

\bibliographystyle{mn2e}

\label{lastpage}
\end{document}